\documentclass[11pt,a4paper]{article}
\usepackage{jheppub}

\usepackage{atlasphysics} 

\usepackage{subfigure}
\usepackage{mathrsfs}
\usepackage{bm}

\usepackage{preprintcover}  
\PreprintCoverPaperTitle{
Search for charged Higgs bosons decaying via $H^{\pm} \rightarrow \tau\nu$ in \\ $t\bar{t}$ events using $pp$ collision data at $\sqrt{s} = 7~\mbox{TeV}$ with \\ the ATLAS detector
}  
\PreprintIdNumber{CERN-PH-EP-2012-083}  
\PreprintCoverAbstract{
The results of a search for charged Higgs bosons are presented.
The analysis is based on \LUMI of proton-proton collision data at
$\sqrt{s}=7$~TeV collected by the ATLAS experiment at the Large Hadron
Collider, using top quark pair events with a $\tau$ lepton in the final
state. The data are consistent with the expected background from Standard
Model processes. Assuming that the branching ratio of the charged Higgs
boson to a $\tau$ lepton and a neutrino is 100\%, this leads to upper
limits on the branching ratio of top quark decays to a $b$ quark and a
charged Higgs boson
between 5\% and 1\% for charged Higgs boson masses ranging from
90~GeV to 160~GeV, respectively. In the context of the $m_h^{\text{max}}$
scenario of the MSSM, $\tan\beta$ above 12--26, as well as between 1
and 2--6, can be excluded for charged Higgs boson masses between 90~GeV
and 150~GeV.
}  
\PreprintJournalName{JHEP}  

\newcommand{\LUMI}{4.6~fb$^{-1}$ }
\def\mt{\ensuremath{m_{\mathrm{T}}}}

\def\jvfcut{\ensuremath{|}JVF\ensuremath{|>0.75}}

\title{\boldmath Search for charged Higgs bosons decaying 
via $H^{\pm} \rightarrow \tau\nu$ in $t\bar{t}$ events 
using $pp$ collision data at $\sqrt{s} = 7~\mbox{TeV}$ 
with the ATLAS detector}

\author{The ATLAS Collaboration}

\date{\today}

\abstract{
The results of a search for charged Higgs bosons are presented. 
The analysis is based on \LUMI of proton-proton collision data at 
$\sqrt{s}=7$~TeV collected by the ATLAS experiment at the Large Hadron 
Collider, using top quark pair events with a $\tau$ lepton in the final 
state. The data are consistent with the expected background from Standard 
Model processes. Assuming that the branching ratio of the charged Higgs 
boson to a $\tau$ lepton and a neutrino is 100\%, this leads to upper 
limits on the branching ratio of top quark decays to a $b$ quark and a 
charged Higgs boson
between 5\% and 1\% for charged Higgs boson masses ranging from 
90~GeV to 160~GeV, respectively. In the context of the $m_h^{\text{max}}$ 
scenario of the MSSM, $\tan\beta$ above 12--26, as well as between 1 
and 2--6, can be excluded for charged Higgs boson masses between 90~GeV 
and 150~GeV.
}

\keywords{Hadron-Hadron Scattering}

\def\mtp2{\ensuremath{m_{\mathrm{T}2}^{H}}}

\begin{document}

\maketitle


\section{Introduction}

Charged Higgs bosons ($H^{+}$, $H^{-}$) are predicted by several 
non-minimal Higgs scenarios, such as Two Higgs Doublet Models 
(2HDM)~\cite{Lee:1973iz} or models containing Higgs 
triplets~\cite{Cheng:1980,Schechter:1980,Lazarides:1981,Mohapatra:1983,Magg:1980}. 
As the Standard Model (SM) does not contain any elementary 
charged scalar particle, the observation of a charged Higgs 
boson\footnote{In the following, charged Higgs bosons are 
denoted $H^+$, with the charge-conjugate $H^-$ always implied. 
Hence, $\tau$ denotes a positively charged $\tau$ lepton.} 
would clearly indicate new physics beyond the SM. 
For instance, supersymmetric models
predict the existence of charged Higgs bosons. In a type-II 2HDM, 
such as the Higgs sector of the Minimal Supersymmetric extension 
of the Standard Model (MSSM)~\cite{Fayet:1976et,Fayet:1977yc,Farrar:1978xj,Fayet:1979sa,Dimopoulos:1981zb}, 
the main $H^+$ production mode at the Large Hadron Collider (LHC) 
is through top quark decays $t \rightarrow bH^+$, for charged 
Higgs boson masses ($m_{H^+}$) smaller than the top quark mass 
($m_{\text{top}}$). The dominant source of top quarks at the LHC 
is through $t\bar{t}$ production. The cross section for $H^+$ 
production from single top quark events is much smaller and is
not considered here. For $\tan\beta > 2$, where $\tan\beta$
is the ratio of the vacuum expectation values of the two 
Higgs doublets, the charged Higgs boson decay via 
$H^{+} \rightarrow \tau\nu$ is dominant and remains 
sizeable for $1 < \tan\beta < 2$~\cite{Dittmaier:2011ti}.
In this paper, ${\cal B}(H^+ \rightarrow \tau\nu) = 100\%$ is assumed, 
unless otherwise specified. Under this assumption, the combined LEP 
lower limit for the charged Higgs boson mass is about 90~GeV~\cite{:2001xy}. 
The Tevatron experiments placed upper limits on 
${\cal B}(t \rightarrow bH^{+})$ in the 15--20\% range for 
$m_{\text{H}^+}<m_{\text{top}}$~\cite{Aaltonen:2009ke,:2009zh}.\\

\begin{figure}[h!]
\begin{center}
\includegraphics[width=0.4\textwidth]{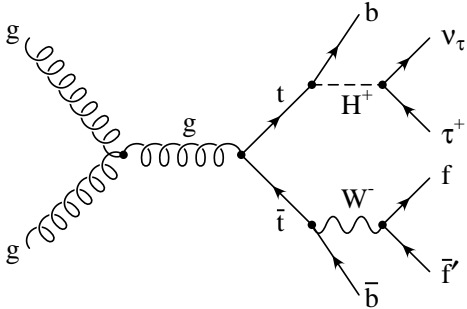}
\end{center}
\vspace*{-3mm}
\caption{Example of a leading-order Feynman diagram for the production 
of $t\bar{t}$ events arising from gluon fusion, where a top quark
decays to a charged Higgs boson, followed by the decay $H^+ \rightarrow \tau\nu$.
\label{fig01}}
\end{figure}

This paper describes a search for charged Higgs bosons with masses 
in the range 90--160~GeV, using $t\bar{t}$ events with a leptonically 
or hadronically decaying $\tau$ lepton in the final state, i.e.\ with the 
topology shown in Fig.~\ref{fig01}. Charged Higgs bosons are 
searched for in a model-independent way, hence exclusion limits  
are given in terms of ${\cal B}(t\rightarrow bH^+)$, as well as in the 
$m_h^{\mathrm{max}}$ scenario~\cite{Carena:2002qg} of the MSSM.
The results are based on \LUMI 
of data from $pp$ collisions at $\sqrt{s}=7~\mbox{TeV}$, collected 
in 2011 with the ATLAS experiment~\cite{Aad:2008zzm} at the LHC.
Three final states, which are expected 
to yield the highest sensitivity, are analysed:
\begin{itemize}
\item lepton+jets: 
$t\bar{t} \to b\bar{b}WH^+ \to b\bar{b}(q\bar{q}')(\tau_{\text{lep}}\nu)$,
i.e.\ $W$ decays hadronically and $\tau$ decays into an electron or a muon, 
with two neutrinos;
\item $\tau$+lepton: 
$t\bar{t} \to b\bar{b}WH^+ \to b\bar{b}(l\nu)(\tau_{\text{had}}\nu)$,
i.e.\ $W$ decays leptonically (with $l = e,\,\mu$) and $\tau$ decays 
hadronically;
\item $\tau$+jets: 
$t\bar{t} \to b\bar{b}WH^+ \to b\bar{b}(q\bar{q}')(\tau_{\text{had}}\nu)$, 
i.e.\ both $W$ and $\tau$ decay hadronically.
\end{itemize}
 
In Section~2, the data and simulated samples used in this analysis are 
described. In Section~3, the reconstruction of physics objects in ATLAS 
is discussed. Sections~4--6 present results obtained in the lepton+jets, 
$\tau$+lepton and $\tau$+jets channels, respectively. Systematic 
uncertainties are discussed in Section~7, before exclusion limits in 
terms of ${\cal B}(t\rightarrow bH^+)$ and $\tan\beta$ are presented 
in Section~8. Finally, a summary is given in Section~9.

\section{Data and simulated events}

The ATLAS detector \cite{Aad:2008zzm} consists of an inner tracking 
detector with a coverage in pseudorapidity\footnote{
ATLAS uses a right-handed coordinate system with
its origin at the nominal interaction point (IP) in the centre
of the detector and the $z$-axis along the beam pipe. The $x$-axis
points from the IP to the centre of the LHC ring, and the $y$-axis
points upwards. Cylindrical coordinates $(r,\phi)$ are used in the
transverse plane, $\phi$ being the azimuthal angle around the beam pipe.
The pseudorapidity is defined in terms of the polar angle $\theta$ as
$\eta=-\ln\tan(\theta/2)$.}
up to $|\eta|=2.5$, surrounded by a thin 2~T superconducting solenoid, 
a calorimeter system extending upto $|\eta|=4.9$ for the detection of 
electrons, photons and hadronic jets, and a large muon spectrometer 
extending up to $|\eta|=2.7$ that measures the deflection of muon 
tracks in the field of three superconducting toroid magnets. A 
three-level trigger system is used. The first level trigger is 
implemented in hardware, using a subset of detector information 
to reduce the event rate to a design value of at most 75~kHz. This 
is followed by two software-based trigger levels, which together 
reduce the event rate to about 300~Hz.\\ 

Only data taken with all ATLAS sub-systems operational 
are used.
It results in an integrated luminosity of \LUMI for the 2011 data-taking 
period. The integrated luminosity has an uncertainty of 3.9\%, measured as 
described in Refs.~\cite{lumi,Aad:2011dr} and based on the whole 2011 
dataset.
Following basic data quality checks, further event cleaning is 
performed by demanding that no jet is consistent with having originated
from instrumental effects, such as large noise signals in one or several 
channels of the hadronic end-cap calorimeter, coherent noise in the 
electromagnetic calorimeter, or non-collision backgrounds. In addition, 
events are discarded if the reconstructed vertex with the largest sum of 
squared track momenta has fewer than five associated tracks with 
transverse momenta $\pt > 400\MeV$.\\

The background processes that enter this search 
include the SM pair production of top quarks 
$t\bar{t} \rightarrow b\bar{b}W^+W^-$, 
as well as the production of single top quark, $W$+jets, $Z/\gamma^*$+jets, 
diboson and multi-jet events. 
Data-driven methods are used in order to estimate the multi-jet 
background, as well as the backgrounds with intrinsic missing transverse 
momentum where electrons or jets are misidentified as hadronically decaying 
$\tau$ leptons.
The modelling of SM $t\bar{t}$ and single top quark events is performed 
with MC@NLO~\cite{Frixione:2002ik}, except for the $t$-channel single 
top quark production where AcerMC~\cite{acer} is used. The top quark 
mass is set to 172.5~GeV and the set of parton distribution functions 
used is CT10~\cite{Lai:2010vv}. For the events generated with MC@NLO, 
the parton shower, hadronisation and underlying event are added using 
HERWIG~\cite{hw} and JIMMY~\cite{jm}. PYTHIA~\cite{Sjostrand:2006za} 
is instead used for events generated with AcerMC. Inclusive cross 
sections are taken from the approximate next-to-next-to-leading-order (NNLO) predictions 
for \ttbar\ production~\cite{hathor}, for single top quark production in the 
$t$-channel and $s$-channel~\cite{Kidonakis:2011wy,Kidonakis:2010tc}, as well 
as for $Wt$ production~\cite{Kidonakis:2010ux}.
Overlaps between $Wt$ and SM $t\bar{t}$ final states are 
removed~\cite{Frixione:2005vw}.
Single vector boson ($W$ and $Z/\gamma^*$) production is 
simulated with ALPGEN~\cite{Mangano:2002ea} interfaced to HERWIG 
and JIMMY, using CTEQ6.1~\cite{Pumplin:2002vw} parton distribution functions. 
The additional partons produced in the matrix element part of the event 
generation can be light partons or heavy quarks. In the latter case, 
dedicated samples with matrix elements for the production of massive
$b\bar{b}$ or $c\bar{c}$ pairs are used. Diboson events ($WW$, $WZ$ 
and $ZZ$) are generated using HERWIG. The cross sections are 
normalised to NNLO predictions for $W$ and $Z/\gamma^*$ 
production~\cite{Gavin:2012kw,Gavin:2010az} and to 
next-to-leading-order (NLO) predictions for diboson 
production~\cite{Campbell:2011bn}.\\

The SM background samples are summarised in 
Table~\ref{xs}. In addition, three types of signal samples are 
produced with PYTHIA for $90\GeV < m_{H^+} < 160\GeV$: 
$t\bar{t} \rightarrow b\bar{b}H^+W^-$, 
$t\bar{t} \rightarrow b\bar{b}H^-W^+$ and 
$t\bar{t} \rightarrow b\bar{b}H^+H^-$, where the 
charged Higgs bosons decay as $H^+ \rightarrow \tau\nu$. 
The cross section for each of these three processes depends 
only on the total $t\bar{t}$ production cross section (167~pb) 
and the branching ratio ${\cal B}(t \to bH^+)$. 
TAUOLA~\cite{Was:2004dg} is used for $\tau$ decays, and 
PHOTOS~\cite{Barberio:1990ms} is used for photon radiation 
from charged leptons.\\

\begin{table}[h!]
  \begin{center}
    \begin{tabular}{l|ll}
Process   &  Generator & Cross section [pb] \\ 
\hline \hline
SM $t\bar{t}$ with at least one lepton $\ell = e,\,\mu,\,\tau$ & 
MC@NLO~\cite{Frixione:2002ik} & 
$91\phantom{.0 \times 10^0}$~\cite{hathor} \\ 
\hline
Single top quark $t$-channel (with $\ell$) & AcerMC~~~\;\cite{acer} &
$21\phantom{.0 \times 10^0}$~\cite{Kidonakis:2011wy} \\ 
Single top quark $s$-channel (with $\ell$) & MC@NLO~\cite{Frixione:2002ik} &
$\phantom{0}1.5\phantom{\times 10^0}$\,\,\,\,\,\cite{Kidonakis:2010tc} \\
Single top quark $Wt$-channel (inclusive) & MC@NLO~\cite{Frixione:2002ik} & 
$16\phantom{.0\times 10^0}$~\cite{Kidonakis:2010ux} \\ 
\hline
$W \rightarrow \ell\nu$ & ALPGEN~~\cite{Mangano:2002ea} & 
$\phantom{0}3.1 \times 10^4$~\cite{Gavin:2012kw} \\ 
\hline
$Z/\gamma^* \rightarrow \ell\ell$ with $m(\ell\ell) > 10~\mbox{GeV}$ & 
ALPGEN~~\cite{Mangano:2002ea} & 
$\phantom{0}1.5 \times 10^4$~\cite{Gavin:2010az} \\
\hline
$WW$  & HERWIG~~\cite{hw} & 
$17\phantom{.0\times 10^0}$~\cite{Campbell:2011bn} \\ 
$ZZ$  & HERWIG~~\cite{hw} & 
$\phantom{0}1.3 \phantom{\times 10^0}$\,\,\,\,\,\cite{Campbell:2011bn} \\
$WZ$  & HERWIG~~\cite{hw} & 
$\phantom{0}5.5 \phantom{\times 10^0}$\,\,\,\,\,\cite{Campbell:2011bn} \\
\hline
$H^+$ signal with ${\cal B}(t \to bH^+)=5\%$ &
PYTHIA~~\,\cite{Sjostrand:2006za} & $16\phantom{.0 \times 10^0}$ \\
    \end{tabular}
\caption{Cross sections for the simulated processes and generators used to model them.
\label{xs}}
  \end{center}
\end{table}

The event generators are tuned in order to describe the ATLAS data. The 
parameter sets AUET2~\cite{herwigtune} and AUET2B~\cite{pythiatune} are 
used for events for which hadronisation is simulated using HERWIG/JIMMY 
and PYTHIA, respectively. 
To take into account the presence of multiple interactions 
(around nine, on average) occurring in the same and neighbouring 
bunch crossings (referred to as pile-up), 
simulated minimum bias events are added to the hard process 
in each generated event. Prior to the analysis, simulated events are 
reweighted in order to match the distribution of the average number of 
pile-up interactions in the data. All generated events are propagated 
through a detailed GEANT4 simulation~\cite{Agostinelli:2002hh,simu} of 
the ATLAS detector and are reconstructed with the same algorithms as the data.


\section{Physics object reconstruction}
\label{sec:object}

\subsection{Electrons}
\label{subsection:electronsec}

Electrons are reconstructed by matching clustered energy deposits 
in the electromagnetic calorimeter to tracks reconstructed in the 
inner detector. The electron candidates are required to meet 
quality requirements based on the expected shower shape~\cite{Aad:2011mk}, 
to have a transverse energy \mbox{$\ET > 20 \GeV$} and to be in 
the fiducial volume of the detector, $|\eta|<2.47$ (the transition 
region between the barrel and end-cap calorimeters, $1.37<|\eta|<1.52$, 
is excluded). Additionally, $\ET$ and $\eta$-dependent calorimeter 
(tracking) isolation requirements are imposed in a cone with a 
radius\footnote{$\Delta R=\sqrt{(\Delta \eta)^2 + (\Delta \phi)^2}$,
where $\Delta \eta$ is the difference in pseudorapidity of the
two objects in question, and $\Delta \phi$ is the difference 
between their azimuthal angles.} $\Delta R = 0.2~(0.3)$ around 
the electron position, excluding the electron object itself, 
with an efficiency of about 90\% for true isolated electrons. 

\subsection{Muons}
\label{subsection:muonsec}

Muon candidates are required to contain matching inner detector and 
muon spectrometer tracks~\cite{sys_muon_rec_eff}, as well as  
to have $\pt > 15 \GeV$ and $|\eta| < 2.5$. 
Only isolated muons are accepted by requiring that 
the transverse energy deposited in the calorimeters 
(the transverse momentum of the inner detector tracks) 
in a cone of radius $\Delta R=0.2$ ($0.3$) around the muon 
amounts to less 
than $4 \GeV$ ($2.5 \GeV$). The energy and momentum of the muon are excluded 
from the cone when applying these isolation requirements. 

\subsection{Jets}

Jets are reconstructed using the anti-$k_t$ 
algorithm~\cite{Cacciari:2008gp, Cacciari:2005hq} 
with a size parameter value of \mbox{$R = 0.4$}. The 
jet finder uses reconstructed three-dimensional, noise-suppressed 
clusters of calorimeter cells~\cite{topoclus}. Jets are calibrated 
to the hadronic energy scale with correction factors based on 
simulation~\cite{Aad:2011he,sys_jes}. A method that allows for the
identification and selection of jets originating from the hard-scatter 
interaction through the use of tracking and vertexing information 
is used~\cite{Abazov:2006yb}. This is referred to as the ``Jet Vertex 
Fraction'' (JVF), defined as the fraction of the total momentum of the 
charged particle tracks associated to the jet which belongs to tracks 
that are also compatible with the primary vertex. By convention, jets 
with no associated tracks are assigned a JVF value of $-1$ in order to 
keep a high efficiency for jets at large values of $\eta$, outside the 
range of the inner tracking detectors. The jet selection based on this 
discriminant is shown to be insensitive to pile-up. A requirement of 
\jvfcut\ is placed on all jets during event selection.
In order to identify the jets initiated by $b$ quarks, an algorithm 
is used that combines impact-parameter information with the explicit 
determination of a secondary vertex~\cite{sys_b}. A working point is 
chosen that corresponds to an average efficiency of about $70\%$ for 
$b$ jets with $\pt>20 \GeV$ in $t\bar{t}$ events and a light-quark jet 
rejection factor of about 130. Since the $b$-tagger relies on the inner 
tracking detectors, the acceptance region for jets is restricted to 
$|\eta| < 2.4$.

\subsection{$\tau$ jets}
\label{subsection:tausec}

In order to reconstruct hadronically decaying $\tau$ leptons, anti-$k_t$ 
jets with either one or three associated tracks reconstructed in the 
inner detector and depositing $\ET > 10 \GeV$ in the calorimeter are 
considered as $\tau$ candidates~\cite{taus}. Dedicated 
algorithms are used in order to reject electrons and muons. Hadronic $\tau$ 
decays are identified using a likelihood criterion designed to 
discriminate against quark- and gluon-initiated jets by using the
shower shape and tracking variables as inputs. A working point  
with an efficiency of about 30\% for hadronically decaying $\tau$ 
leptons with $\pt > 20 \GeV$ 
in $Z \rightarrow \tau\tau$ events is chosen, leading to a rejection factor 
of about 100--1000 
for
jets. The rejection factor depends on the
$\pt$ and $\eta$ of 
the candidate and the number of associated tracks. The $\tau$ candidates 
are further required to have a visible transverse momentum of at 
least 20~GeV and to be within $|\eta| < 2.3$. The selected $\tau$ candidates 
are henceforth referred to as ``$\tau$ jets''.

\subsection{Removal of geometric overlaps between objects}
\label{subsection:overlap}

When candidates selected using the criteria above overlap geometrically, 
the following procedures are applied, in this order: muon candidates are 
rejected if they are found within $\Delta R < 0.4$ of any jet with 
$\pt > 25 \GeV$; a $\tau$ jet is rejected if found 
within $\Delta R < 0.2$ of a selected muon or electron; jets are removed 
if they are within $\Delta R < 0.2$ of a selected $\tau$ object or electron.

\subsection{Missing transverse momentum}

The missing transverse momentum and its magnitude $\met$~\cite{Aad:2012re} 
are reconstructed from three-dimensional, noise-suppressed clusters of 
cells in the calorimeter and from muon tracks reconstructed in the muon 
spectrometer and the inner tracking detectors. Clusters of calorimeter 
cells belonging to jets (including $\tau$ jets) with $\pt > 20 \GeV$ 
are calibrated to the hadronic energy scale. 
Calorimeter cells not associated with any object are also taken into 
account and they are calibrated at the electromagnetic energy scale. 
In order to deal appropriately with the energy deposited by muons in 
the calorimeters, the contributions of muons to $\met$ are calculated 
differently for isolated and non-isolated muons.


\section{Analysis of the lepton+jets channel}
\label{sec:lepjets}

This analysis relies on the detection of lepton+jets 
decays of $t\bar{t}$ events, where the charged lepton $l$ 
(electron or muon) arises from $H^+ \rightarrow \tau_{\text{lep}}\nu$, 
while the jets arise from a hadronically decaying $W$ boson, i.e.\ 
$t\bar{t} \to b\bar{b}WH^+ \to b\bar{b}(q\bar{q}')(\tau_{\text{lep}}\nu)$.

\subsection{Event selection}

The lepton+jets analysis uses events passing a single-lepton 
trigger with an $\et$ threshold of 20--22~GeV for electrons\footnote{
The electron trigger threshold was increased from 20~GeV to 22~GeV towards the 
end of data-taking in 2011.} and a $\pt$ threshold of 18~GeV for muons. 
These thresholds are low enough to guarantee that electrons and muons 
chosen for the analysis are in the plateau region of the trigger-efficiency 
curve. In addition, to select a sample of lepton+jets events enriched in 
$t\bar{t}$ candidates, the following requirements are applied:
\begin{itemize}
  \item exactly one lepton having $E_{\text{T}} > 25\GeV$ 
(electron) or $\pt > 20\GeV$ (muon) and matched to the 
corresponding trigger object, with neither a second lepton nor a 
$\tau$ jet in the event;
  \item at least four jets having $\pt > 20\GeV$, with exactly two 
of them being $b$-tagged;
  \item $\met > 40~\mbox{GeV}$ and, in order to discriminate between 
$\met$ arising from isolated neutrinos and from poorly reconstructed 
leptons, this requirement is tightened to 
$\met \times |\sin\Delta\phi_{l,\text{miss}}| > 20~\mbox{GeV}$ 
if the azimuthal angle $\Delta\phi_{l,\text{miss}}$ between the 
lepton and $\met$ is smaller than $\pi/6$. 
\end{itemize}

Having selected a lepton+jets sample enriched in $t\bar{t}$ 
candidates, jets must be assigned correctly to the decay products 
of each $W$ boson (with a mass $m_W = 80.4~\mbox{GeV}$) and top quark. 
In particular, the hadronic side of the event is identified by selecting 
the combination of one $b$-tagged jet ($b$) and two untagged jets ($j$) 
that minimises:
\begin{equation}
\label{jl_eq_chi2}
\chi^2 =
\frac{ (m_{jjb} - m_{\text{top}})^2}{\sigma_{\text{top}}^2}
+ \frac{(m_{jj} - m_{W})^2}{\sigma_W^2},
\end{equation}
where $\sigma_{\text{top}}=17~\mbox{GeV}$ and $\sigma_W=10~\mbox{GeV}$ 
are the widths of the reconstructed top quark and $W$ boson mass 
distributions, as measured in simulated $t\bar{t}$ events. 
Using information about 
the correctly identified combinations in the generated events, the 
jet assignment efficiency is found to be 72\%. Events with $\chi^2 > 5$ 
are rejected in order to select well-reconstructed hadronic top 
quark candidates.

\subsection{Data-driven estimation of backgrounds with 
misidentified leptons}
\label{subsection:fakeleptongeneral}

While the ATLAS lepton identification gives a very pure sample of 
candidates, there is a non-negligible contribution from non-isolated 
leptons arising from the semileptonic decay of hadrons containing $b$ 
or $c$ quarks, 
from the decay-in-flight of $\pi^{\pm}$ or $K$ mesons and, in the 
case of misidentified electron objects, from the reconstruction of  
$\pi^{0}$ mesons, photon conversions or shower fluctuations. All leptons 
coming from such mechanisms are referred to as \emph{misidentified} 
leptons, as opposed to truly isolated leptons (e.g.\ from the prompt 
decay of $W$ or $Z$ bosons), which are referred to as \emph{real} leptons. 
The data-driven estimation of the number of misidentified leptons passing 
the lepton selections of Sections~\ref{subsection:electronsec} 
and~\ref{subsection:muonsec} is based on exploiting differences in 
the lepton identification between real and misidentified electrons 
or muons. Two data samples are defined, which differ only in the lepton 
identification criteria. The \emph{tight} sample contains mostly events 
with real leptons and uses the same lepton selection as in the analysis. 
The \emph{loose} sample contains mostly events with misidentified leptons. 
This latter sample is obtained by loosening the isolation and identification 
requirements for the leptons. 
For loose electrons, the isolation requirements have an efficiency of 
about 98\% for true isolated electrons, compared to 90\% in the tight 
sample. For loose muons, the isolation requirement is removed.
By construction, the tight sample is therefore 
a subset of the loose sample.\\

Let $N_{\text{r}}^{\text{L}}$ and $N_{\text{m}}^{\text{L}}$ 
($N_{\text{r}}^{\text{T}}$ and $N_{\text{m}}^{\text{T}}$) 
be the number of events containing real and misidentified leptons, 
respectively, passing a loose (tight) selection. The numbers of events 
containing one loose or tight lepton are given by:    
\begin{eqnarray} 
N^{\text{L}} & = & N_{\text{m}}^{\text{L}}+N_{\text{r}}^{\text{L}}, \\
N^{\text{T}} & = & N_{\text{m}}^{\text{T}}+N_{\text{r}}^{\text{T}}. 
\end{eqnarray} 
Defining $p_{\text{r}}$ and $p_{\text{m}}$ as:
\begin{equation} 
p_{\text{r}} = \frac{N^{\text{T}}_{\text{r}}}{N^{\text{L}}_{\text{r}}} \quad 
\text{and} 
\quad p_{\text{m}} = \frac{N^{\text{T}}_{\text{m}}}{N^{\text{L}}_{\text{m}}},
\end{equation} 
the number of misidentified leptons passing the tight selection 
$N_{\text{m}}^{\text{T}}$ can then be written as: 
\begin{equation} 
N_{\text{m}}^{\text{T}} = 
\dfrac{p_{\text{m}}}{p_{\text{r}}-p_{\text{m}}}(p_{\text{r}}N^{\text{L}} - N^{\text{T}}). 
\end{equation} 

The main ingredients of this data-driven method are thus the relative 
efficiencies $p_{\text{r}}$ and $p_{\text{m}}$ for a real or a 
misidentified lepton, respectively, to be detected as a tight lepton. 
The lepton identification efficiency $p_{\text{r}}$ is measured using 
a tag-and-probe method on $Z \rightarrow ll$ data events with a dilepton 
invariant mass between 86~GeV and 96~GeV, where one lepton is required to 
fulfill tight selection criteria. The rate at which the other lepton 
passes the same tight selection criteria defines $p_{\text{r}}$. The 
average values of the electron and muon identification efficiencies 
are 80\% and 97\%, respectively. 
On the other hand, a control sample with misidentified leptons is selected 
by considering events in the data with exactly one lepton passing the loose 
criteria. In order to select events dominated by multi-jet production, 
$\met$ is required to be between 5~GeV and 20~GeV. Residual true leptons 
contribute at a level below 10\% and are subtracted from this sample using 
simulation. After this subtraction, the rate at which a loose lepton passes 
tight selection criteria defines the misidentification rate $p_{\text{m}}$. 
The average values of the electron and muon misidentification probabilities 
are 18\% and 29\%, respectively. In the final parameterisation of 
$p_{\text{r}}$ and $p_{\text{m}}$, dependencies on the pseudorapidity 
of the lepton, its distance $\Delta R$ to the nearest jet and the leading 
jet $\pt$ are taken into account.

\subsection{Reconstruction of discriminating variables after the selection cuts}

The analysis uses two variables that discriminate between leptons 
produced in $\tau \rightarrow l\nu_l\nu_{\tau}$ and leptons coming 
directly from $W$ boson decays. 
The first discriminating variable is the invariant mass $m_{bl}$ of 
the $b$ jet and the charged lepton $l$ coming from the same top quark 
candidate, or more conveniently, $\cos\theta_{l}^*$ defined as:
\begin{equation}
\label{eq2}
\cos\theta_{l}^* = \frac{2m_{bl}^2}{m_{\text{top}}^2-m_W^2}-1 \simeq 
\frac{4\,p^b \cdot p^l}{m_{\text{top}}^2-m_W^2}-1.
\end{equation}
Both $m_b^2$ and $m_l^2$ are neglected, hence 
$m_{bl}^2 \simeq 2\,p^b \cdot p^l$, where $p^b$ 
and $p^{l}$ are the four-momenta of the $b$ jet and 
of the charged lepton $l$, respectively. The presence 
of a charged Higgs boson in a leptonic top quark decay 
reduces the invariant product $p^b \cdot p^{l}$, when compared 
to $W$-mediated top quark decays, leading to $\cos\theta_{l}^*$ 
values closer to $-1$.\\

The second discriminating variable is the transverse mass 
$m_{\mathrm{T}}^H$~\cite{Gross:2009wia}, obtained by fulfilling the constraint 
$(p^{\text{miss}} + p^l + p^b)^2  = m_{\text{top}}^2$ on the 
leptonic side of lepton+jets $\ttbar$ events. More than one neutrino 
accounts for the invisible four-momentum $p^{\text{miss}}$ and its transverse 
component $\vec \pt^{\text{miss}}$. By construction, $m_{\mathrm{T}}^H$ 
gives an event-by-event lower bound on the mass of the leptonically 
decaying charged ($W$ or Higgs) boson produced in the top quark decay, 
and it can be written as: 
\begin{equation}
(m_{\mathrm{T}}^H )^2  = 
\left(\sqrt {m_{\text{top}}^2  + 
(\vec \pt^{l} + \vec \pt^{b} + 
\vec \pt^{\text{miss}})^2} - \pt^{b} \right)^2 - 
\left(\vec \pt^{l} + \vec \pt^{\text{miss}}\right)^2.
\label{eq:mth2}
\end{equation}

\begin{figure}[h!]
\vspace*{-4mm}
\centering
\subfigure[]{
\includegraphics[width=0.45\textwidth]{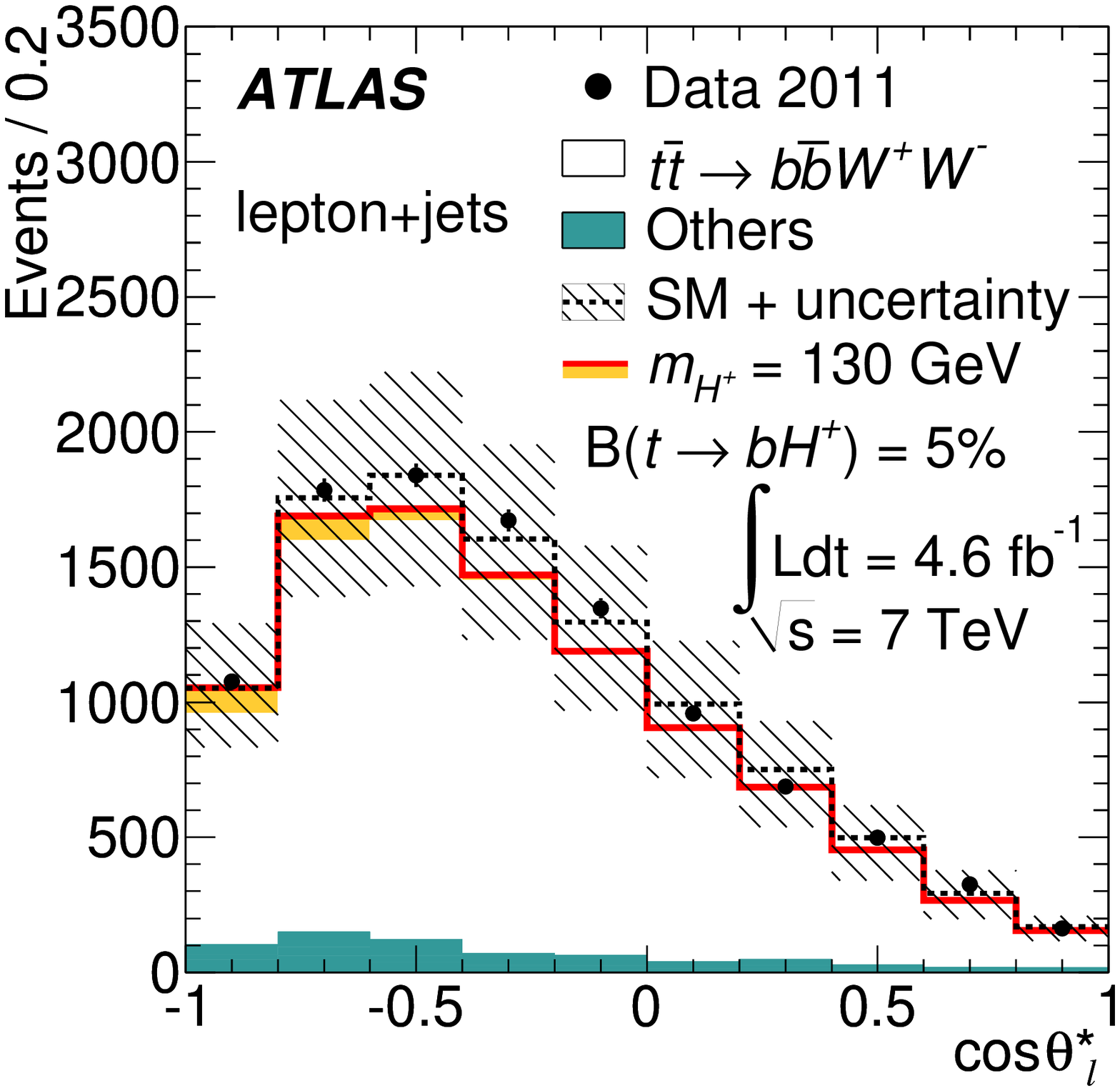}
\label{one-costheta-mth:subfig1}
}
\subfigure[]{
\includegraphics[width=0.45\textwidth]{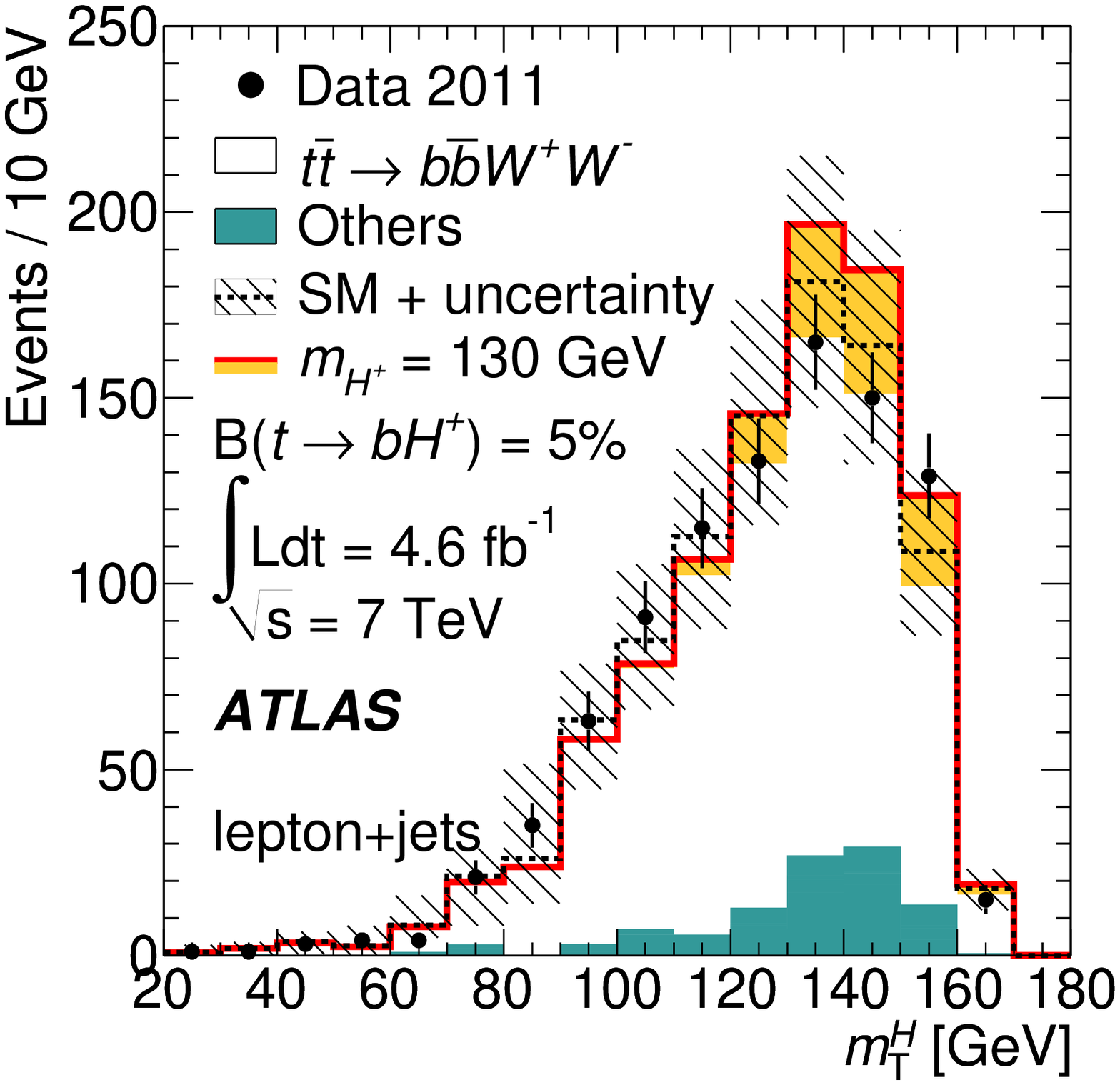}
\label{one-costheta-mth:subfig2}
}
\caption{Distribution of (a) $\cos\theta_l^*$  and (b) $m_{\mathrm{T}}^H$, 
in the signal region ($\cos\theta_l^* < -0.6$, 
$m_{\mathrm{T}}^W < 60~\mbox{GeV}$) for the 
latter. The dashed line corresponds to the SM-only hypothesis and the 
hatched area around it shows the total uncertainty for the SM backgrounds, 
where ``Others'' refers to the contribution of all SM processes 
except $t\bar{t} \rightarrow b\bar{b}W^+W^-$. The solid line shows the 
predicted contribution of signal+background in the presence of a 130~GeV 
charged Higgs boson, assuming 
${\cal B}(t \rightarrow bH^{+}) = 5\%$ and 
${\cal B}(H^+ \rightarrow \tau\nu) = 100\%$. The light area below the 
solid line corresponds to the contribution of the $H^+$ signal, stacked on 
top of the scaled $t\bar{t} \rightarrow b\bar{b}W^+W^-$ background and 
other SM processes.
\label{one-costheta-mth}}
\end{figure}

The $\cos\theta_l^*$ distribution measured in the data is shown 
in Fig.~\ref{one-costheta-mth:subfig1}, superimposed on the predicted 
background, determined with a data-driven method for the multi-jet background 
and simulation for the other SM backgrounds. In the presence of a 
charged Higgs boson in the top quark decays, with a branching ratio 
${\cal B}(t\rightarrow bH^+)$, the contribution of  
$t\bar{t} \rightarrow b\bar{b}W^+W^-$ events in the background is 
scaled according to this branching ratio. A control region enriched 
in $t\bar{t} \rightarrow b\bar{b}W^+W^-$ events is defined by requiring 
$-0.2 < \cos\theta_l^* < 1$. In Section~\ref{sec:limit}, this sample is 
used to fit the branching ratio ${\cal B}(t \rightarrow bH^{+})$ and the 
product of the cross section $\sigma_{bbWW}$, the luminosity, the selection 
efficiency and acceptance for $t\bar{t} \rightarrow b\bar{b}W^+W^-$, 
simultaneously with the likelihood for the signal estimation. In turn, 
this ensures that the final results, and in particular the upper limit 
on ${\cal B}(t \rightarrow bH^{+})$, are independent of the 
assumed theoretical production cross section for $t\bar{t}$. With 
a branching fraction ${\cal B}(t\rightarrow bH^+)=5\%$, the signal 
contamination in the control region would range from 1.3\% for 
$m_{H^+} = 90~\mbox{GeV}$ to 0.4\% for $m_{H^+} = 160~\mbox{GeV}$.
The signal region is defined by requiring 
$\cos\theta_l^* < -0.6$ and $m_{\mathrm{T}}^W < 60~\mbox{GeV}$, where:
\begin{equation}
m_{\mathrm{T}}^W = 
\sqrt{2 \pt^l  \met (1 - \cos\Delta\phi_{l,\text{miss}})}.
\end{equation}
This is done in order to suppress the background from events with 
a $W$ boson decaying directly into electrons or muons. 
For events in the signal region, $m_{\mathrm{T}}^H$, shown in 
Fig.~\ref{one-costheta-mth:subfig2}, is used as a discriminating 
variable to search for charged Higgs bosons. 
Table~\ref{tab:results_lepjets} lists the contributions to the signal 
region of the SM processes and of $t\bar{t}$ events with at least one 
decay $t \rightarrow bH^+$, assuming $m_{H^+} = 130~\mbox{GeV}$ and 
${\cal B}(t\rightarrow bH^+)=5\%$. When including signal 
in the prediction, the simulated SM $t\bar{t}$ contribution is scaled 
according to this branching ratio.
The data are consistent with the predicted SM background 
and no significant deformation of the $m_{\mathrm{T}}^H$ distribution 
is observed.
 
\begin{table}[h!]
\begin{center}
\begin{tabular}{l|l} 
Sample & Event yield (lepton+jets)\\
\hline \hline
$t\bar t$ & $840\phantom{.0}\pm20\phantom{.0}\pm150\phantom{.0}$  \\
Single top quark & $\phantom{0}28\phantom{.0}\pm\phantom{0}2\phantom{.00}^{+8}_{-6}$  \\
$W$+jets & $\phantom{0}14\phantom{.0}\pm\phantom{0}3\phantom{.00}^{+6}_{-3}$ \\
$Z$+jets & $\phantom{00}2.1\pm\phantom{0}0.7\phantom{0}^{+1.2}_{-0.4}$ \\
Diboson & $\phantom{00}0.5\pm\phantom{0}0.1\pm\phantom{00}0.2$ \\
Misidentified leptons & $\phantom{0}55\phantom{.0}\pm10\phantom{.0}\pm\phantom{0}20\phantom{.0}$ \\
\hline
All SM backgrounds & $940\phantom{.0}\pm22\phantom{.0} \pm150\phantom{.0}$ \\
\hline
Data & 933 \\
\hline
$t \rightarrow bH^+$ (130~GeV) & $120\phantom{.0}\pm\phantom{0}4\phantom{.0}\pm\phantom{0}25\phantom{.0}$ \\
Signal+background & $990\phantom{.0}\pm21\phantom{.0} \pm140\phantom{.0}$\\
\end{tabular}
\caption{Expected event yields in the signal region 
of the lepton+jets final state, and comparison with \LUMI of data. 
A cross section of 167~pb is assumed for the SM $t\bar{t}$ background.
The numbers shown in the last two rows, for a 
hypothetical $H^+$ signal with $m_{H^+} = 130~\mbox{GeV}$, are obtained 
with ${\cal B}(t \rightarrow bH^{+}) = 5\%$. Both statistical and 
systematic uncertainties are shown, in this order.
}
\label{tab:results_lepjets}
\end{center}
\end{table}

\section{Analysis of the $\tau$+lepton channel}
\label{sec:taulep}

This analysis relies on the detection of $\tau$+lepton decays of 
$t\bar{t}$ events, where the hadronically decaying $\tau$ lepton
arises from $H^+ \rightarrow \tau_{\text{had}}\nu$, while an 
electron or muon comes from the decay of the $W$ boson, i.e.\ 
$t\bar{t} \to b\bar{b}WH^+ \to b\bar{b}(l\nu)(\tau_{\text{had}}\nu)$.

\subsection{Event selection}
The $\tau$+lepton analysis relies on the same single-lepton trigger 
signatures as the lepton+jets analysis presented in Section~\ref{sec:lepjets}.
In order to select $\tau$+lepton 
events, the following requirements are made:
\begin{itemize}
  \item exactly one lepton, having $E_{\text{T}} > 25~\mbox{GeV}$ 
(electron) or $\pt > 20~\mbox{GeV}$ (muon) and matched to the 
corresponding trigger object, and no other electron or muon;
  \item exactly one $\tau$ jet having $\pt > 20~\mbox{GeV}$ and an electric 
charge opposite to that of the lepton;
  \item at least two jets having $\pt > 20~\mbox{GeV}$, including at 
least one $b$-tagged jet;
  \item $\sum \pt > 100~\mbox{GeV}$ in order to suppress multi-jet events, 
where $\sum \pt$ is the sum of the transverse momenta of all tracks 
associated with the primary vertex. Tracks entering the sum must pass 
quality cuts on the number of hits and have $\pt > 1~\mbox{GeV}$. As 
this variable is based on tracks from the primary vertex (as opposed 
to energy deposits in the calorimeter), it is robust against pile-up.
\end{itemize}

$\met$ is used as the discriminating variable to distinguish 
between SM $t\bar{t}$ events and those where top quark decays are 
mediated by a charged Higgs boson, in which case the neutrinos are 
likely to carry away more energy. 

\subsection{Data-driven estimation of backgrounds with misidentified leptons}
\label{sec:qcd_taulep}
The estimation of the backgrounds with misidentified leptons uses the 
data-driven method described in Section~\ref{subsection:fakeleptongeneral}. 
When implementing the method, the dependence of real and misidentification 
rates on the $b$-tagged jet multiplicity are taken into account, as well as 
the requirement for one $\tau$ jet (instead of a $\tau$ jet veto).

\subsection{Backgrounds with electrons and jets misidentified as $\tau$ jets}
\label{sec:taufakes_taulep}
The background with electrons misidentified as $\tau$ jets is estimated 
using a $Z \to ee$ control region in the data~\cite{taus}, where one 
electron is reconstructed as a $\tau$ jet. The measured misidentification 
probabilities, which have an average value of 0.2\%, are then applied to 
all simulated events in the $\tau$+lepton analysis. Simulation studies 
show that this application is valid, as the misidentification probabilities 
for $Z \to ee$ and $t\bar{t}$ events are similar.\\

A data-driven method applied to a control sample enriched in $W$+jets 
events is used to measure the probability for a jet to be 
misidentified as a hadronically decaying $\tau$ lepton. This measured 
probability is used to predict the yield of background events due 
to jet$\,\to\tau$ misidentification. Like jets from the hard process 
in the dominant \ttbar\ background, jets in the control sample originate 
predominantly from quarks instead of gluons. The main difference between 
\ttbar\ and $W$+jets events is the different fraction of $b$ jets, which 
is smaller in $W$+jets events. However, the probability for a $b$ jet 
to be misidentified as a $\tau$ jet is smaller than the corresponding 
probability for a light-quark jet, because the average track multiplicity 
is higher for $b$ jets. Moreover, the visible mass measurement used in 
the $\tau$ identification provides further discrimination between $b$ 
jets and $\tau$ jets. Differences in jet composition (e.g.\ the ratio 
of gluons to quarks) between \ttbar\ and $W$+jets, assessed using 
simulation, are taken into account as systematic uncertainties. 
These also cover the dependence of the probability on whether a 
$b$ jet or a light-quark jet is misidentified as a $\tau$ jet.
Events in the control region are required to pass the same single-lepton 
trigger, data quality and lepton requirements as in the $\tau$+lepton 
event selection. Additionally, a $\tau$ candidate and $\met>40~\mbox{GeV}$ 
are required, and events with $b$-tagged jets are vetoed. Simulated events 
with a true $\tau$ contribute at a level below 0.5\% and are subtracted.
The $\tau$ candidates are required to have $\pt > 20~\mbox{GeV}$, 
$\abseta < 2.3$, and cannot be within $\Delta R=0.2$ of any electron 
or muon. They are also not required to pass $\tau$ identification. The 
jet$\,\to\tau$ misidentification probability is defined as the number 
of objects passing the full $\tau$ identification divided by the number 
prior to requiring identification.
This misidentification probability is evaluated separately for $\tau$ 
candidates with one or three associated tracks (the corresponding average 
values are about 7\% and 2\%, respectively) and, in addition, it is 
measured as a function of both $\pt$ and $\eta$.\\
 
In order to predict the background for the charged Higgs boson search, 
the measured jet$\,\to\tau$ misidentification probability is applied to 
simulated \ttbar, single top quark, $W$+jets, $Z/\gamma^*$+jets and diboson 
events, all of which are required to pass the full event selection except 
for the $\tau$ identification. 
For these events, $\tau$ candidates not overlapping with a true $\tau$ 
lepton or a true electron, but otherwise fulfilling the same requirements 
as in the denominator of the misidentification probability, are identified.
Each of them is considered separately to be potentially misidentified 
as a $\tau$ jet. In order to avoid counting the same object twice, each 
jet that corresponds to a $\tau$ candidate is removed from the event. 
The number of reconstructed jets and the number of $b$-tagged jets are 
adjusted accordingly. If, after taking this into consideration, the event 
passes the $\tau$+lepton selection, it is counted as a background event 
with a weight given by the misidentification probability corresponding to 
the $\pt$ and $\eta$ of the $\tau$ candidate. The predicted numbers 
of events from this data-driven method and from simulation 
are shown in Table~\ref{tab:fake_rate_applied}.
The backgrounds arising from the jet$\,\to\tau$ misidentification are not 
well modelled in simulation, which is why they are estimated using 
data-driven methods.

\begin{table}[h!]
\begin{center}
\begin{tabular}{l|c|c}
\centering
Sample & Data-driven method [events] & Simulation [events] \\
\hline \hline
$\ttbar$         &  $900 \pm                   15$  & $877          \pm 6$ \\
$W$+jets         &  $150 \pm \phantom0 3$           & $145          \pm 9$ \\
Single top quark & $\phantom0 81 \pm \phantom0 1$   & $\phantom0 61 \pm 2$ \\
$Z/\gamma^*$+jets& $\phantom0 44 \pm \phantom0 1$   & $\phantom0 69 \pm 4$ \\
Diboson          & $\phantom{00} 6 \pm \phantom0 1$ & $\phantom{00}8\pm 1$\\
\end{tabular}
\caption{Application of the misidentification probability obtained 
from $W$+jets events in the data, for the
$\tau$+lepton channel. The predictions of the background 
contributions based on data-driven misidentification probabilities 
and on simulation are given, with statistical uncertainties 
only. In both cases, all top quarks are assumed to decay via 
$t \rightarrow bW$. 
}
\label{tab:fake_rate_applied}
\end{center}
\end{table}

\subsection{Event yields and $\met$ distribution after the selection cuts}
\begin{figure}[h!]
\centering
\subfigure[]{
\includegraphics[width=0.45\textwidth]{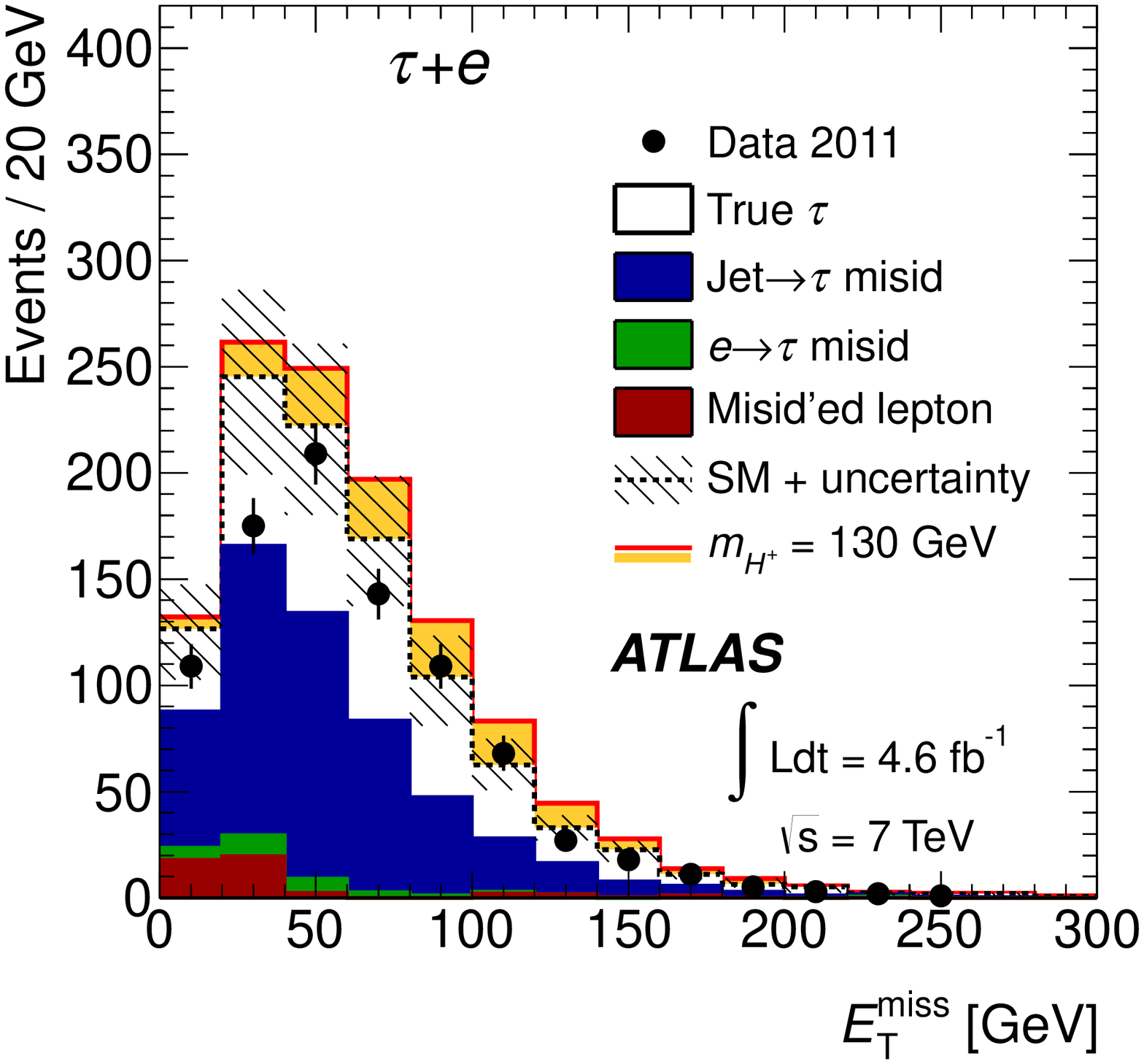}
\label{fig:results_taulep:subfig1}
}
\subfigure[]{
\includegraphics[width=0.45\textwidth]{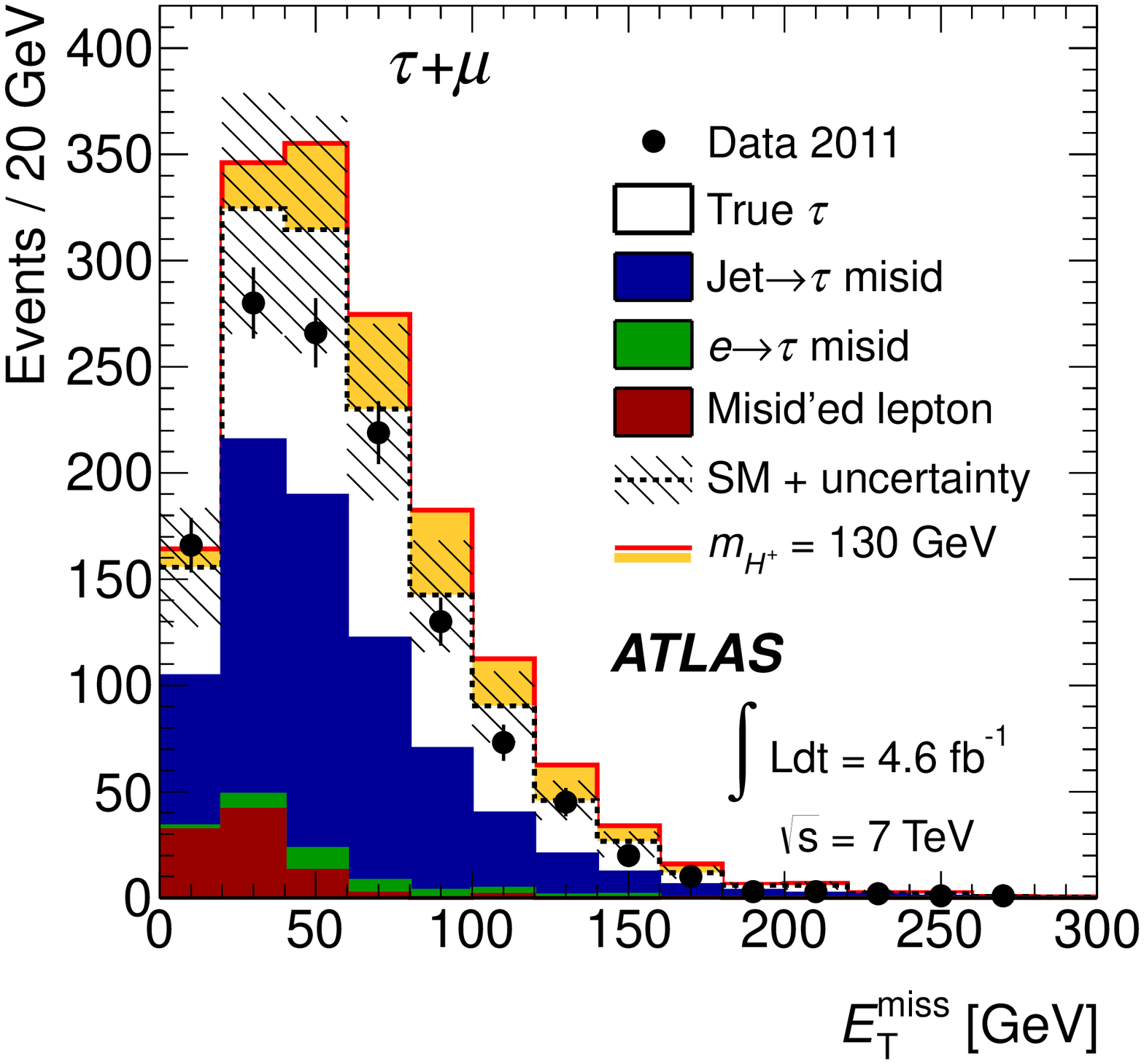}
\label{fig:results_taulep:subfig2}
}
\caption{
$\met$ distribution after all selection cuts in the $\tau$+lepton channel, 
for (a) $\tau$+electron and (b) $\tau$+muon final states. The dashed line 
corresponds to the SM-only hypothesis and the hatched area around it shows 
the total uncertainty for the SM backgrounds. The solid line shows the 
predicted contribution of signal+background in the presence of a 
130~GeV charged Higgs boson with 
${\cal B}(t \rightarrow bH^{+}) = 5\%$ and 
${\cal B}(H^+ \rightarrow \tau\nu) = 100\%$. The 
contributions of $t\bar{t} \rightarrow b\bar{b}W^+W^-$ events 
in the backgrounds with true or misidentified $\tau$ jets are 
scaled down accordingly.
\label{fig:results_taulep}}
\end{figure}

Table~\ref{tab:results_taulep} shows the expected number of 
background events for the SM-only hypothesis and the observation 
in the data. The total number of predicted 
events (signal+background) in the presence of a 130~GeV charged Higgs 
boson with ${\cal B}(t \rightarrow bH^{+}) = 5\%$ is also shown. 
The $\tau$+lepton analysis relies on the 
theoretical \ttbar\ production cross section 
$\sigma_{t\bar{t}} = 167^{+17}_{-18}~\mbox{pb}$~\cite{hathor} 
for the background estimation. In the presence of a charged 
Higgs boson in the top quark decays, with a branching ratio 
${\cal B}(t\rightarrow bH^+)$, the contributions of  
$t\bar{t} \rightarrow b\bar{b}W^+W^-$ events in the backgrounds with 
true or misidentified $\tau$ jets are scaled 
according to this branching ratio. 
The background 
with correctly reconstructed $\tau$ jets is obtained with simulation. 
The data are found to be consistent with the expectation for the 
background-only hypothesis. The $\met$ distributions for the $\tau+e$ 
and $\tau+\mu$ channels, after all selection cuts are applied, are shown in 
Fig.~\ref{fig:results_taulep}.

\begin{table}[h!]
\begin{center}
\begin{tabular}{l|ll} 
Sample & \multicolumn{2}{c}{Event yield ($\tau$+lepton)} \\
       & \multicolumn{1}{c}{$\tau+e$} 
       & \multicolumn{1}{c}{$\tau+\mu$}   \\
\hline \hline
True $\tau$+lepton & $\phantom{0}430 \pm 14\pm\phantom{0}59$ 
                   & $\phantom{0}570 \pm 15\pm\phantom{0}75$ \\
Misidentified jet$\,\to\tau$ & $\phantom{0}510 \pm 23\pm\phantom{0}86$
             & $\phantom{0}660 \pm 26\pm110$\\
Misidentified $e\to\tau$   & $\phantom{00}33 \pm \phantom{0}4\pm\phantom{00}5$
             & $\phantom{00}34 \pm \phantom{0}4\pm\phantom{00}6$ \\
Misidentified leptons & $\phantom{00}39 \pm10\pm\phantom{0}20$ 
             & $\phantom{00}90 \pm10\pm\phantom{0}34$\\
\hline
All SM backgrounds   & $          1010 \pm30\pm110$
             & $          1360 \pm30\pm140$ \\
\hline
Data         &  $\phantom{0}880$ & 1219 \\
\hline
$t \rightarrow bH^+$ (130~GeV) 
& $\phantom{0}220\pm\phantom{0}6\pm\phantom{0}29$
& $\phantom{0}310\pm\phantom{0}7\pm\phantom{0}39$ \\
Signal+background &  $1160 \pm30\pm100$  
& $1570 \pm30\pm130$\\
\end{tabular}
\caption{Expected event yields after all selection cuts in the
$\tau$+lepton channel and comparison with \LUMI of data. The numbers 
in the last two rows, obtained for a hypothetical $H^+$ signal 
with $m_{H^+} = 130~\mbox{GeV}$, are obtained with
${\cal B}(t \rightarrow bH^{+}) = 5\%$. All other rows 
assume ${\cal B}(t \rightarrow bW) = 100\%$.
Both statistical and 
systematic uncertainties are shown, in this order.}
\label{tab:results_taulep}
\end{center}
\end{table}

\section{Analysis of the $\tau$+jets channel}\label{sec:taujets}

The analysis presented here relies on the detection of $\tau$+jets 
decays of $t\bar{t}$ events, where the hadronically decaying $\tau$ 
lepton arises from $H^+ \rightarrow \tau_{\text{had}}\nu$, while the 
jets come from a hadronically decaying $W$ boson, i.e.\ 
$t\bar{t} \to b\bar{b}WH^+ \to b\bar{b}(q\bar{q}')(\tau_{\text{had}}\nu)$.

\subsection{Event selection}
\label{sec:event_taujets}

The $\tau$+jets analysis uses events passing a $\tau+\met$ trigger with 
a threshold of 29~GeV on the $\tau$ object and 35~GeV on calorimeter-based 
$\met$.
The following requirements are applied, in this order:
\begin{itemize}
  \item at least four jets (excluding $\tau$ jets) having $\pt > 20~\GeV$, 
of which at least one is $b$-tagged;
  \item exactly one $\tau$ jet with $\pt^\tau > 40~\GeV$, found 
within $|\eta|<2.3$ and matched to a $\tau$ trigger object; 
  \item neither a second $\tau$ jet with $\pt^\tau > 20~\GeV$, 
nor any electrons with $\et > 20~\GeV$, nor any muons with $\pt > 15~\GeV$;
  \item $\met > 65~\mbox{GeV}$;
  \item to reject events in which a large reconstructed $\met$ 
is due to the limited resolution of the energy measurement, the 
following ratio based on the $\sum\pT$ definition of 
Section~\ref{sec:taulep} must satisfy:
\[
\frac{\met}{0.5\GeV^{1/2} \cdot \sqrt{\sum\pT}} > 13;
\] 
  \item a topology consistent with a top quark decay: the 
combination of one $b$-tagged jet ($b$) and two untagged jets ($j$) with 
the highest $\pt^{jjb}$ must satisfy $m_{jjb} \in [120, 240] \GeV$.
\end{itemize}

For the selected events, the transverse mass $\mt$ 
is defined as:
\begin{equation}
\mt=\sqrt{ 2 \pt^\tau  \ET^{\mathrm{miss}} (1-\cos \Delta\phi_{\tau,\text{miss}}) },
\end{equation}
where $\Delta\phi_{\tau,\text{miss}}$ is the azimuthal angle between the 
$\tau$ jet and the direction of the missing momentum. This discriminating 
variable is related to the $W$ boson mass in the $W\rightarrow \tau\nu$ 
background case and to the $H^+$ mass for the signal hypothesis.

\subsection{Data-driven estimation of the multi-jet background}
\label{sec:qcd_taujets}
\begin{figure}[h!] 
\centering   
\subfigure[]{ \includegraphics[width=0.45\textwidth]{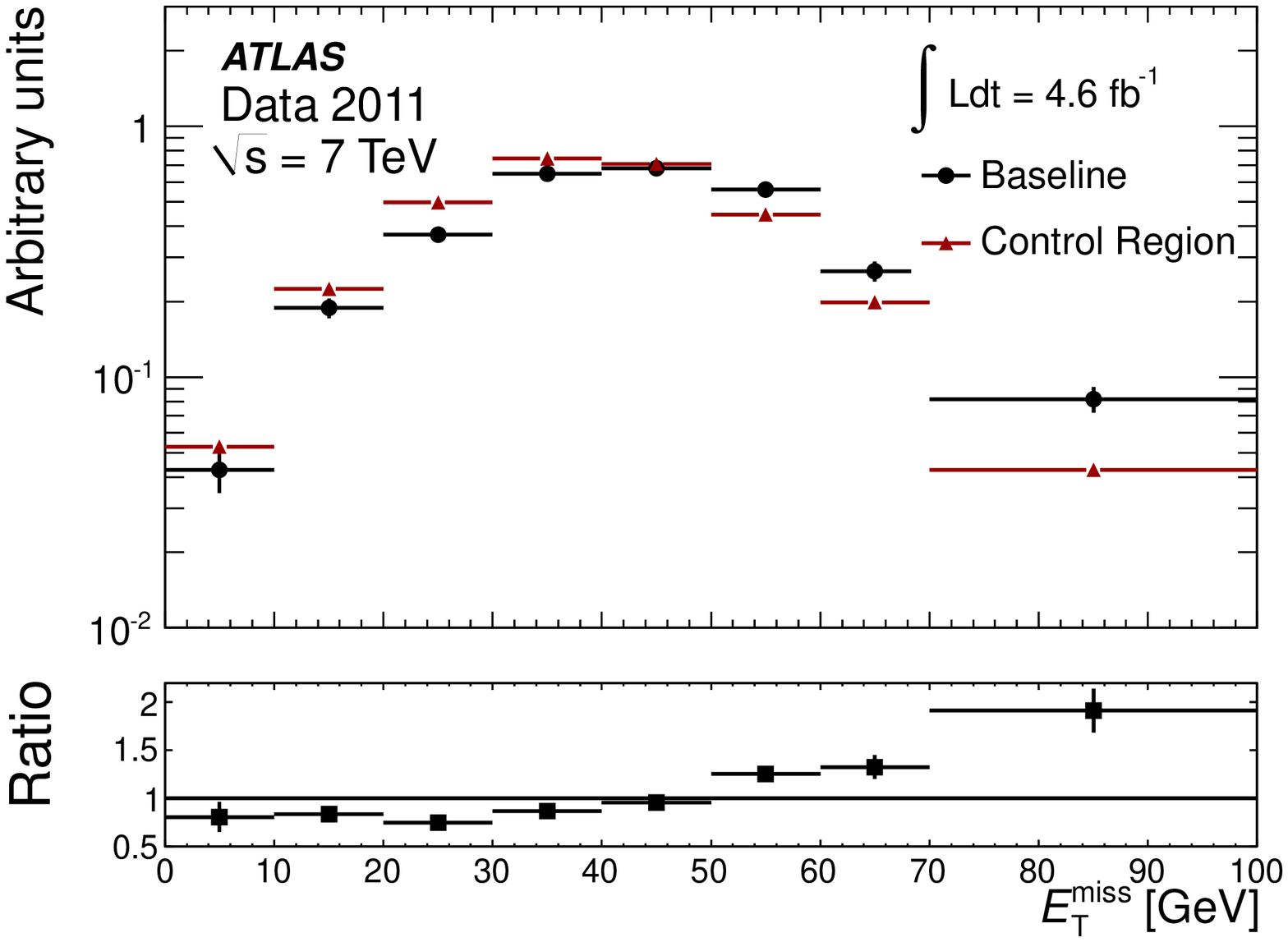} 
\label{fig:qcd_compare:subfig1}
}
\subfigure[]{ \includegraphics[width=0.45\textwidth]{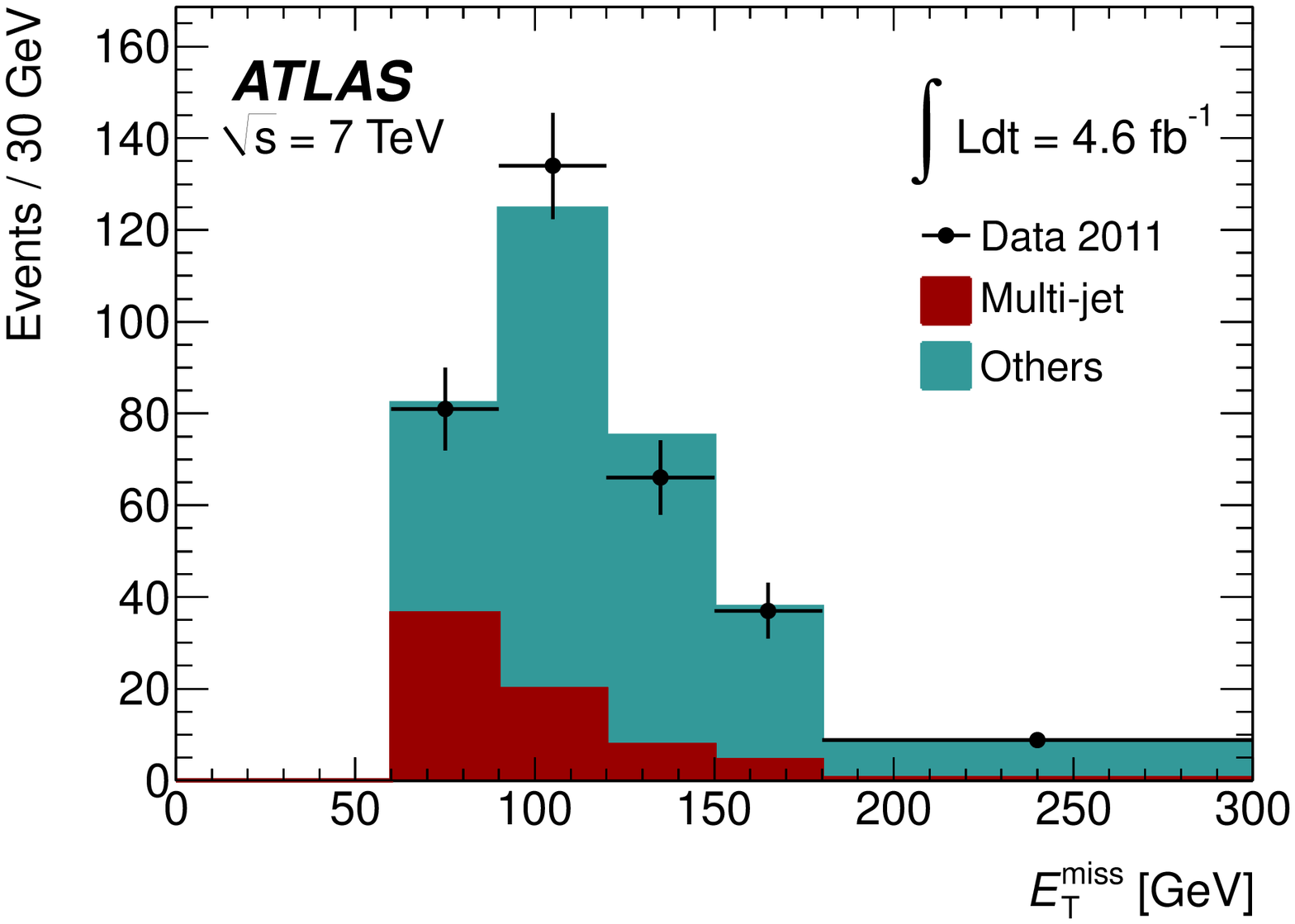}  
\label{fig:qcd_compare:subfig2}
}
\caption{
(a) Shape of $\met$ in a control region of the data or using the baseline 
selection, after subtracting the expectation from \ttbar, $W$+jets, and 
single top quark processes estimated from simulation. The distributions are 
compared just before the $\met$ requirement in the baseline selection 
of Section~\ref{sec:event_taujets}, with the exception that, in 
the control region, the $\tau$ selection and the $b$-tagging 
requirements are modified, see text. 
(b) Fit of the $\met$ template to data, in the signal region. Only 
statistical uncertainties are shown.
}
\label{fig:qcd_compare}
\end{figure}

The multi-jet background is estimated by fitting its $\met$ shape 
(and the $\met$ shape of other backgrounds) to data. In order to study this 
shape in a data-driven way, a control region is defined where the 
$\tau$ identification and $b$-tagging requirements are modified, 
i.e.\ $\tau$ candidates must pass a loose $\tau$ identification but 
fail the tight $\tau$ identification used in the signal selection, 
and the event is required not to contain any $b$-tagged jet. Hence, 
the requirement on $m_{jjb}$ is also removed. Assuming that the 
shapes of the $\met$ and $\mt$ distributions are the same in the 
control and signal regions, the $\met$ shape for the multi-jet 
background is measured in the control region, after subtracting 
the simulated background contributions from other processes. These 
other processes amount to less than $1\%$ of the observed events 
in the control region. 
The $\met$ shapes obtained with the $\tau$+jets selection of 
Section~\ref{sec:event_taujets} or in the control region are 
compared just before the $\met$ requirement in the baseline 
selection in Fig.~\ref{fig:qcd_compare:subfig1}. The differences 
between the two distributions are accounted for as systematic 
uncertainties. For the baseline selection, the $\met$ distribution 
measured in the data is then fit using two shapes: the multi-jet model 
and the sum of other processes (dominated by $\ttbar$ and $W$+jets), 
for which the shape and the relative normalisation are taken from 
simulation, as shown in Fig.~\ref{fig:qcd_compare:subfig2}. The ratio 
between the numbers of multi-jet background events in the control and 
signal regions enters the likelihood function for the signal estimation 
(see Section~\ref{sec:limit}) as a nuisance parameter while the shape of 
the multi-jet background is measured in the same region after additionally 
requiring $\met>65\GeV$.

\subsection{Backgrounds with electrons and jets misidentified as $\tau$ jets}
\label{sec:taufakes_taujets}

The methods described in Section~\ref{sec:taufakes_taulep} are used to 
estimate the probability for electrons or jets to be misidentified as 
$\tau$ jets. The estimated contribution to the background from the 
jet$\,\to\tau$ misidentification after the $\tau$+jets selection is given 
in Table~\ref{tab:jetfake_rate_applied_taujets}. 
The backgrounds arising from the jet$\,\to\tau$ misidentification are not 
expected to be well modelled in simulation, which is why they are estimated 
using data-driven methods.

\begin{table}[h!]
\begin{center}
\begin{tabular}{l|c|c}
\centering
Sample     & Data-driven method [events]  & 
Simulation  [events] \\
\hline \hline
$\ttbar$   &  $33\phantom{.0} \pm 1\phantom{.0}$  & $37\phantom{.0} \pm 1\phantom{.0}$ \\
$W$+jets & $\phantom{0}2.5 \pm 0.1$  & 
           $\phantom{0}3.9 \pm 1.5$ \\
Single top quark & $\phantom{0}1.3 \pm 0.1$  &  
                   $\phantom{0}2.0 \pm 0.3$ \\
\end{tabular}
\caption{Application of the misidentification probability obtained 
from a control region in the data enriched in $W$+jets events, for the 
$\tau$+jets channel. The predictions of the background contributions 
based on data-driven misidentification probabilities and on simulation 
are given, with statistical uncertainties only. In both cases, all top 
quarks decay via $t \rightarrow bW$. 
}
\label{tab:jetfake_rate_applied_taujets}
\end{center}
\end{table}

\subsection{Data-driven estimation of backgrounds with correctly 
reconstructed $\tau$ jets}
\label{sec:embedding_taujets}

An embedding method~\cite{Aad:2011rv} 
is used to estimate the backgrounds that contain correctly 
reconstructed $\tau$ jets. The method consists of selecting a control sample 
of $t\bar{t}$-like $\mu$+jets events and replacing the detector signature 
of the muon by a simulated hadronic $\tau$ decay. These new hybrid events 
are then used for the background prediction.
In order to select this
control sample from the data, the following event selection is applied:
\begin{itemize}
\item event triggered by a single-muon trigger with a $\pt$ threshold of 
$18\GeV$;
\item exactly one isolated muon with $\pt >25\GeV$, no isolated electron with 
$\ET >20\GeV$;
\item at least four jets with $\pt >20\GeV$, at least one of which is $b$-tagged;
\item $\ET^{\mathrm{miss}}>35\GeV$.
\end{itemize}

This selection is looser than the selection defined in 
Section~\ref{sec:event_taujets} in order not to bias the control sample. 
The impurity from the background with muons produced in $\tau$ decays 
and non-isolated muons (dominantly $b\bar{b}$ and $c\bar{c}$ events) 
is about $10\%$. However, this contribution is greatly reduced 
as these events are much less likely to pass the $\tau$+jets selection, 
in particular the $\pt^{\tau}$ requirement.\\

\begin{figure}[h!]
\begin{center}
\includegraphics[width=0.57\textwidth]{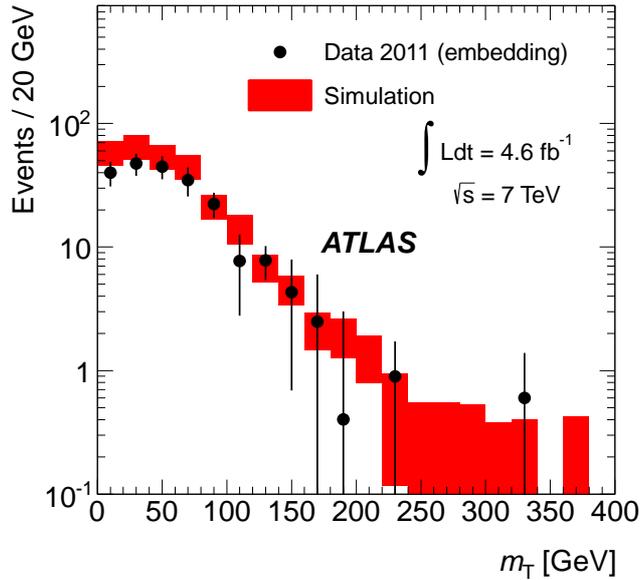}
\end{center}
\caption{
Comparison of the $m_{\text{T}}$ distribution for correctly reconstructed 
$\tau$ jets, predicted by the embedding method and simulation. 
Combined statistical and systematic uncertainties (as described in 
Section~\ref{sec:systematics}) are shown.
\label{fig:taujets_emb}}
\end{figure}

The shape of the $\mt$ distribution for the backgrounds with true $\tau$ 
jets is taken from the distribution obtained with the embedded events, 
after having applied the $\tau$+jets event selection. The 
normalisation is then derived from the number of embedded events: 
\begin{equation}
N_{\tau} = N_{\mathrm{embedded}} \cdot \left( 1 - c_{\tau\to\mu} \right) 
\frac{\epsilon^{\tau+\met-\mathrm{trigger}}}{\epsilon^{\mu-\mathrm{ID, trigger}}} \cdot {\cal B}(\tau \rightarrow \mathrm{hadrons}+\nu),
\end{equation}
where $N_\tau$ is the estimated number of events with correctly 
reconstructed $\tau$ jets, $N_{\mathrm{embedded}}$ is the number 
of embedded events in the signal region, $c_{\tau\to\mu}$ is the 
fraction of events in which the selected muon is a decay product 
of a $\tau$ lepton (taken from simulation), 
$\epsilon^{\tau+\met-\mathrm{trigger}}$ is the 
$\tau + \met$ trigger efficiency (as a function of 
$p_T^\tau$ and $\met$, derived from data), 
$\epsilon^{\mu-\mathrm{ID, trigger}}$ is the muon trigger and identification 
efficiency (as a function of $p_T$ and $\eta$, derived from data) and 
${\cal B}(\tau \rightarrow \mathrm{hadrons}+\nu)$ is the branching ratio of 
the $\tau$ lepton decays involving hadrons.
The $m_{\text{T}}$ distribution for correctly reconstructed $\tau$ jets, 
as predicted by the embedding method, is shown in Fig.~\ref{fig:taujets_emb} 
and compared to simulation.\\
	
\subsection{Event yields and $m_{\text{T}}$ distribution after the selection cuts}

\begin{figure}[h!]
\begin{center}
\includegraphics[width=0.57\textwidth]{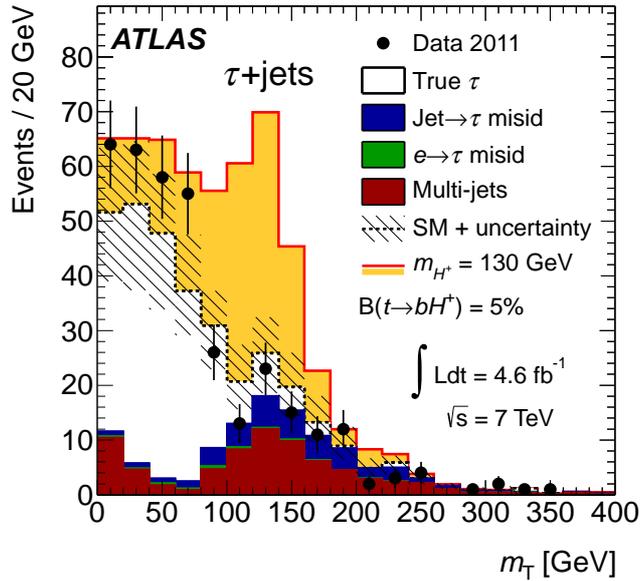}
\end{center}
\caption{
Distribution of $m_{\text{T}}$ after all selection cuts in the $\tau$+jets 
channel. The dashed line corresponds to the SM-only hypothesis and the 
hatched area around it shows the total uncertainty for the SM backgrounds. 
The solid line shows the predicted contribution of signal+background 
in the presence of a charged Higgs boson with $m_{H^+} = 130~\mbox{GeV}$, 
assuming ${\cal B}(t \rightarrow bH^{+}) = 5\%$ and 
${\cal B}(H^+ \rightarrow \tau\nu) = 100\%$. The 
contributions of $t\bar{t} \rightarrow b\bar{b}W^+W^-$ events 
in the backgrounds with misidentified objects are 
scaled down accordingly.
\label{fig:results_taujets}}
\end{figure}

Table~\ref{tab:results_taujets} shows the expected number of 
background events for the SM-only hypothesis and the 
observation in the data. The total number of predicted events 
(signal+background) in the presence of a 130~GeV charged Higgs 
boson with ${\cal B}(t \rightarrow bH^{+}) = 5\%$ is also shown.
The number of events with a correctly reconstructed $\tau$ jet is derived 
from the number of embedded events and does not depend on the 
cross section of the $t\bar{t} \rightarrow b\bar{b}W^+W^-$ process. 
On the other hand, the $\tau$+jets analysis relies on the 
theoretical inclusive \ttbar\ production cross section 
$\sigma_{t\bar{t}} = 167^{+17}_{-18}~\mbox{pb}$~\cite{hathor}
for the estimation of the background with electrons or 
jets misidentified as $\tau$ jets. In the presence of a charged 
Higgs boson in the top quark decays, with a branching ratio 
${\cal B}(t\rightarrow bH^+)$, the contributions of 
$t\bar{t} \rightarrow b\bar{b}W^+W^-$ events in these backgrounds are 
scaled according to this branching ratio. 
The data are found to be consistent with the 
estimation of the SM background. The $\mt$ distribution for the 
$\tau$+jets channel, after all selection cuts are applied, is 
shown in Fig.~\ref{fig:results_taujets}.\\

\begin{table}[h!]
\begin{center}
\begin{tabular}{l|l}
Sample & Event yield ($\tau$+jets) \\
\hline \hline
True $\tau$ (embedding method) & $210\pm10\pm44$\\
Misidentified jet$\,\to\tau$ & $\phantom{0}36\pm\phantom{0}6\pm10$\\
Misidentified $e\to\tau$   & $\phantom{00}3\pm\phantom{0}1\pm\phantom{0}1$\\
Multi-jet processes & $\phantom{0}74\pm\phantom{0}3\pm47$\\
\hline
All SM backgrounds   & $330\pm12\pm65$ \\
\hline
Data         &  $355$                   \\
\hline
$t \rightarrow bH^+$ (130~GeV) & $220\pm\phantom{0}6\pm56$ \\
Signal+background & $540\pm13\pm85$        \\
\end{tabular}
\caption{Expected event yields after all selection cuts in the
$\tau$+jets channel and comparison with \LUMI of data. The 
numbers in the last two rows, obtained for a hypothetical $H^+$ 
signal with $m_{H^+} = 130~\mbox{GeV}$, are obtained with
${\cal B}(t \rightarrow bH^{+}) = 5\%$. 
The rows for the backgrounds with misidentified 
objects assume ${\cal B}(t \rightarrow bW) = 100\%$.
Both statistical 
and systematic uncertainties are shown, in this order.
\label{tab:results_taujets}
}
\end{center}
\end{table}

\section{Systematic uncertainties}
\label{sec:systematics}

\subsection{Systematic uncertainties arising from the detector simulation}

Systematic uncertainties arising from 
the simulation of pile-up and object reconstruction  
are considered. The latter arise from the simulation of the 
trigger, from the reconstruction and identification efficiencies, 
as well as from the energy/momentum scale and resolution for the 
objects described in Section~\ref{sec:object}. To assess the impact 
of most sources of systematic uncertainty, the selection 
cuts for each analysis are re-applied after shifting a particular 
parameter by its $\pm 1$ standard deviation uncertainty. The 
systematic uncertainties related to the electrons and muons 
are discussed in, respectively, Ref.~\cite{Aad:2011mk} and
Refs.~\cite{sys_muon_rec_eff,sys_muon_mom_res}. For the jets, 
see Ref.~\cite{sys_jes} and, in particular, Ref.~\cite{sys_b} for 
the $b$-tagging calibration. The systematic uncertainties related 
to $\tau$ jets are discussed in Ref.~\cite{taus}. Finally, for 
the reconstruction of $\met$, see Ref.~\cite{Aad:2012re}. All studies 
of systematic uncertainties have been updated with the full dataset 
collected in 2011.
The dominant instrumental systematic uncertainties arise from the jet 
energy resolution (10--30\%, depending on $\pt$ and $\eta$), the jet 
energy scale (up to 14\%, depending on $\pt$ and $\eta$, to which a 
pile-up term of 2--7\% and a $b$ jet term of 2.5\% are added in 
quadrature), as well as the $b$-tagging efficiency (5--17\%, 
depending on $\pt$ and $\eta$) and misidentification probability 
(12--21\%, depending on $\pt$ and $\eta$). In comparison, the systematic 
uncertainties arising from the reconstruction and identification 
of electrons and muons are small. All instrumental systematic 
uncertainties are also propagated to the reconstructed $\met$.

\subsection{Systematic uncertainties arising from the generation of \ttbar\ events}

In order to estimate the systematic uncertainties arising from the
$t\bar t$ generation and the parton shower model, the acceptance
is computed for $t\bar t$ events produced with MC@NLO
interfaced to HERWIG/JIMMY and POWHEG~\cite{Frixione:2007vw} 
interfaced to PYTHIA. For the signal samples, which are generated with
PYTHIA (i.e.\ without higher-order corrections), no alternative generator is
available. Instead, the systematic uncertainty for the signal samples
is set to the relative difference in acceptance between $t\bar t$ events
generated with MC@NLO interfaced to HERWIG/JIMMY and with
AcerMC, which is also a leading-order generator,
interfaced to PYTHIA. The systematic uncertainties arising from
initial and final state radiation are computed using $t\bar{t}$ samples
generated with AcerMC interfaced to PYTHIA, where initial and final state
radiation parameters are set to a range of values not excluded by the 
experimental data~\cite{isrfsr}. The largest relative differences with 
respect to the reference sample after full event selections are used as 
systematic uncertainties. 
The systematic uncertainties arising from the modelling of the $t\bar{t}$ 
event generation and the parton shower, as well as initial and final state 
radiation, are summarised in Table~\ref{tab:sysgen} for each analysis.

\begin{table}[h!]
\begin{center}
{
\begin{tabular}{l|r}
Source of uncertainty & Normalisation uncertainty \\
\hline\hline
lepton+jets: \\
\hline
\; Generator and parton shower ($b\bar{b}WH^+$, signal region)  & 10\%  \\
\; Generator and parton shower ($b\bar{b}W^+W^-$, signal region)  & 8\%  \\
\; Generator and parton shower ($b\bar{b}WH^+$, control region)  &  7\% \\
\; Generator and parton shower ($b\bar{b}W^+W^-$, control region)  &  6\% \\
\; Initial and final state radiation  (signal region) &  8\% \\
\; Initial and final state radiation  (control region) &  13\% \\
\hline\hline
$\tau$+lepton: \\
\hline
\; Generator and parton shower ($b\bar{b}WH^+$)  & 2\%  \\
\; Generator and parton shower ($b\bar{b}W^+W^-$)  & 5\%  \\
\; Initial and final state radiation  & 13\%  \\
\hline\hline
$\tau$+jets: \\
\hline
\; Generator and parton shower ($b\bar{b}WH^+$)  & 5\%  \\
\; Generator and parton shower ($b\bar{b}W^+W^-$)  & 5\%  \\
\; Initial and final state radiation  & 19\%  \\
\end{tabular}
}
\end{center}
\caption{\label{tab:sysgen}
Systematic uncertainties arising from the modelling of 
$\ttbar \to b\bar{b}W^+W^-$ and $\ttbar \to b\bar{b}WH^+$ events
and the parton shower, as well as from initial and final state radiation.
}
\end{table}
\subsection{Systematic uncertainties arising from data-driven background estimates}

The systematic uncertainties arising from the data-driven 
methods used to estimate the various backgrounds are summarised 
in Table~\ref{tab:systmethod}, for each of the three channels 
considered in the analysis.\\

\begin{table}
\begin{center}
{
\begin{tabular}{l|r|r}
Source of uncertainty & Normalisation uncertainty & Shape uncertainty \\
\hline\hline
lepton+jets: lepton misidentification \\
\hline
\; Choice of control region       & $6\%$  & - \\
\; $Z$ mass window                & $4\%$  & - \\
\; Jet energy scale               & $16\%$  & - \\
\; Jet energy resolution          & $7\%$  & - \\
\; Sample composition             & $31\%$  & - \\
\hline\hline
$\tau$+lepton: jet$\,\to\tau$ misidentification \\
\hline
\; Statistics in control region   & $2\%$  & - \\
\; Jet composition                & $11\%$ & - \\
\; Object-related systematics     & $23\%$ & $3\%$ \\
\hline
$\tau$+lepton: $e\to\tau$ misidentification \\
\hline
\; Misidentification probability  & $20\%$  & - \\
\hline
$\tau$+lepton: lepton misidentification \\
\hline
\; Choice of control region       & $ 4\%$  & - \\
\; $Z$ mass window                & $ 5\%$  & - \\
\; Jet energy scale               & $14\%$  & - \\
\; Jet energy resolution          & $ 4\%$  & - \\
\; Sample composition             & $39\%$  & - \\
\hline\hline
$\tau$+jets: true $\tau$ \\
\hline
\; Embedding parameters           & $6\%$        & $3\%$ \\
\; Muon isolation                 & $7\%$        & $2\%$\\
\; Parameters in normalisation    & $16\%$       & - \\
\; $\tau$ identification          & $5\%$        & - \\
\; $\tau$ energy scale            & $6\%$        & $1\%$ \\
\hline
$\tau$+jets: jet$\,\to\tau$ misidentification \\
\hline
\; Statistics in control region   & $2\%$  & - \\
\; Jet composition                & $12\%$ & - \\
\; Purity in control region       & $6\%$  & $1\%$ \\
\; Object-related systematics     & $21\%$ & $2\%$ \\
\hline
$\tau$+jets: $e\to\tau$ misidentification \\
\hline
\; Misidentification probability  & $22\%$  & - \\
\hline
$\tau$+jets: multi-jet estimate \\
\hline
\; Fit-related uncertainties      & $32\%$ & - \\
\; $\met$-shape in control region & $16\%$ & - \\
\end{tabular}
}
\end{center}
\caption{\label{tab:systmethod}
Dominant systematic uncertainties on the data-driven estimates. The shape 
uncertainty given is the relative shift of the mean value of the final 
discriminant distribution. A ``-'' in the second column indicates 
negligible shape uncertainties.
\vspace*{2mm}
}
\end{table}

For backgrounds with misidentified leptons, 
discussed in Sections~\ref{subsection:fakeleptongeneral} 
and~\ref{sec:qcd_taulep}, the main systematic uncertainties 
arise from the simulated samples used for subtracting 
true leptons in the determination of the misidentification 
probabilities. These are sensitive to the instrumental systematic 
uncertainties and to the sample dependence (misidentification 
probabilities are calculated in a control region dominated 
by gluon-initiated events, but later used in a data sample 
with a higher fraction of quark-initiated events).\\

The dominant systematic uncertainties in the estimation of the 
multi-jet background in the $\tau$+jets channel, described in 
Section~\ref{sec:qcd_taujets}, are the statistical uncertainty 
of the fit due to the limited size of the data control sample 
and uncertainties due to potential differences of the $\met$ 
shape in the signal and control regions. 
The dominant systematic uncertainties in estimating the contribution 
of events with electrons misidentified as $\tau$ jets in 
Sections~\ref{sec:taufakes_taulep} and~\ref{sec:taufakes_taujets} 
arise from the subtraction of the multi-jet and electroweak 
backgrounds in the control region enriched with $Z \rightarrow ee$ 
events and from potential correlations in the selections of the tag 
and probe electrons. For the estimation of backgrounds with jets 
misidentified as hadronically decaying $\tau$ leptons, also discussed 
in Sections~\ref{sec:taufakes_taulep} and~\ref{sec:taufakes_taujets}, 
the dominant systematic uncertainties on the misidentification 
probability are the statistical uncertainty due to the limited 
control sample size and uncertainties due to the difference of 
the jet composition (gluon or quark-initiated) in the control 
and signal regions, which is estimated using simulation.
Other uncertainties come from the impurities arising from 
multi-jet background events and from true hadronic $\tau$ decays 
in the control sample. The systematic uncertainties affecting the 
estimation of the background 
from correctly reconstructed $\tau$ jets in the $\tau$+jets channel,  
discussed in Section~\ref{sec:embedding_taujets}, consist of the potential 
bias introduced by the embedding method itself, uncertainties from 
the trigger efficiency measurement, uncertainties associated to 
simulated $\tau$ jets ($\tau$ energy scale and identification 
efficiency) and uncertainties on the normalisation, which 
are dominated by the statistical uncertainty of the selected control 
sample and the $\tau+\met$ trigger efficiency uncertainties.


\section{Results}
\label{sec:limit}
\begin{figure}[t!h!]
  \centering
\subfigure[]{
    \includegraphics[width=0.47\textwidth]{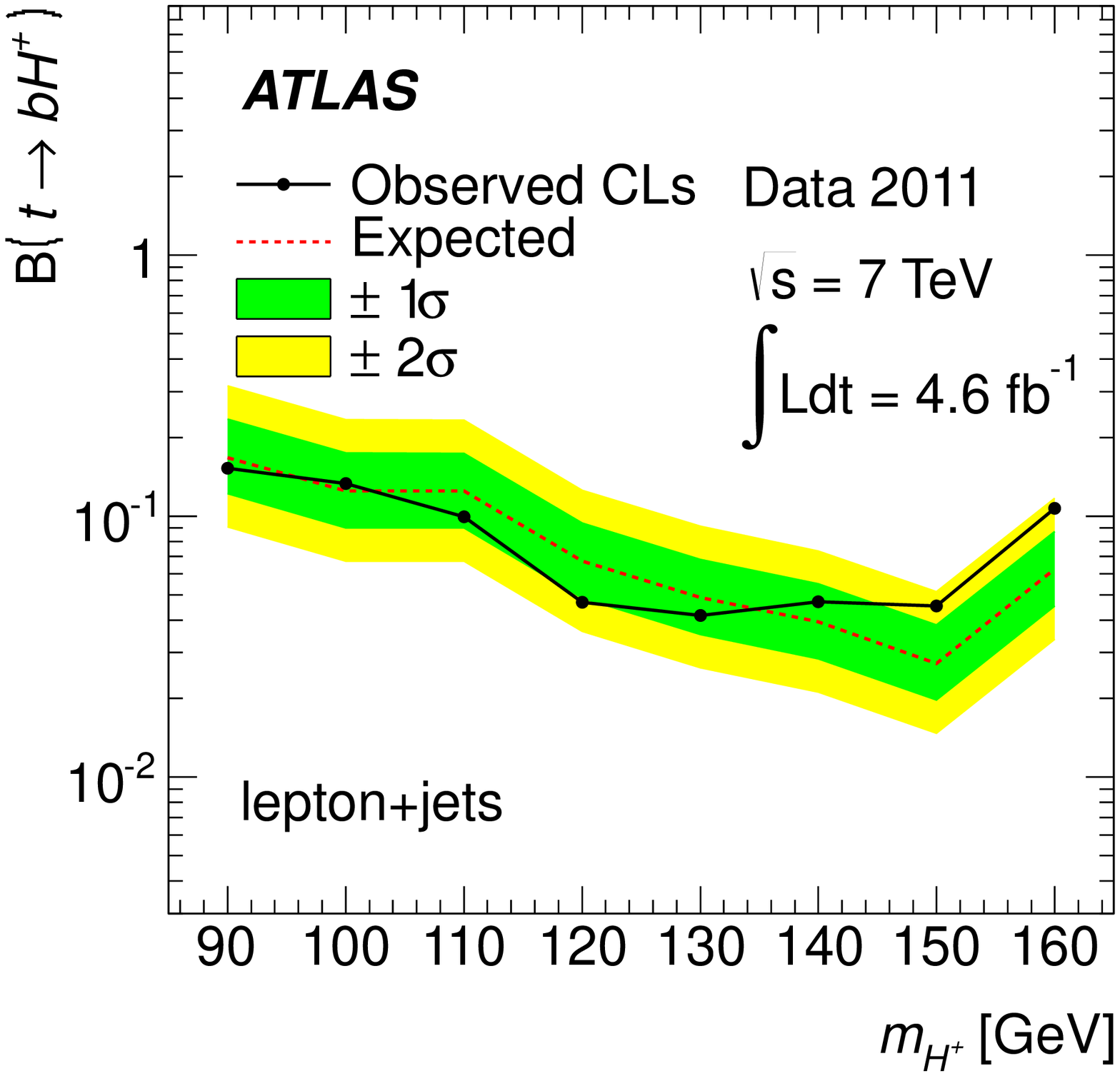}
}
\subfigure[]{
    \includegraphics[width=0.47\textwidth]{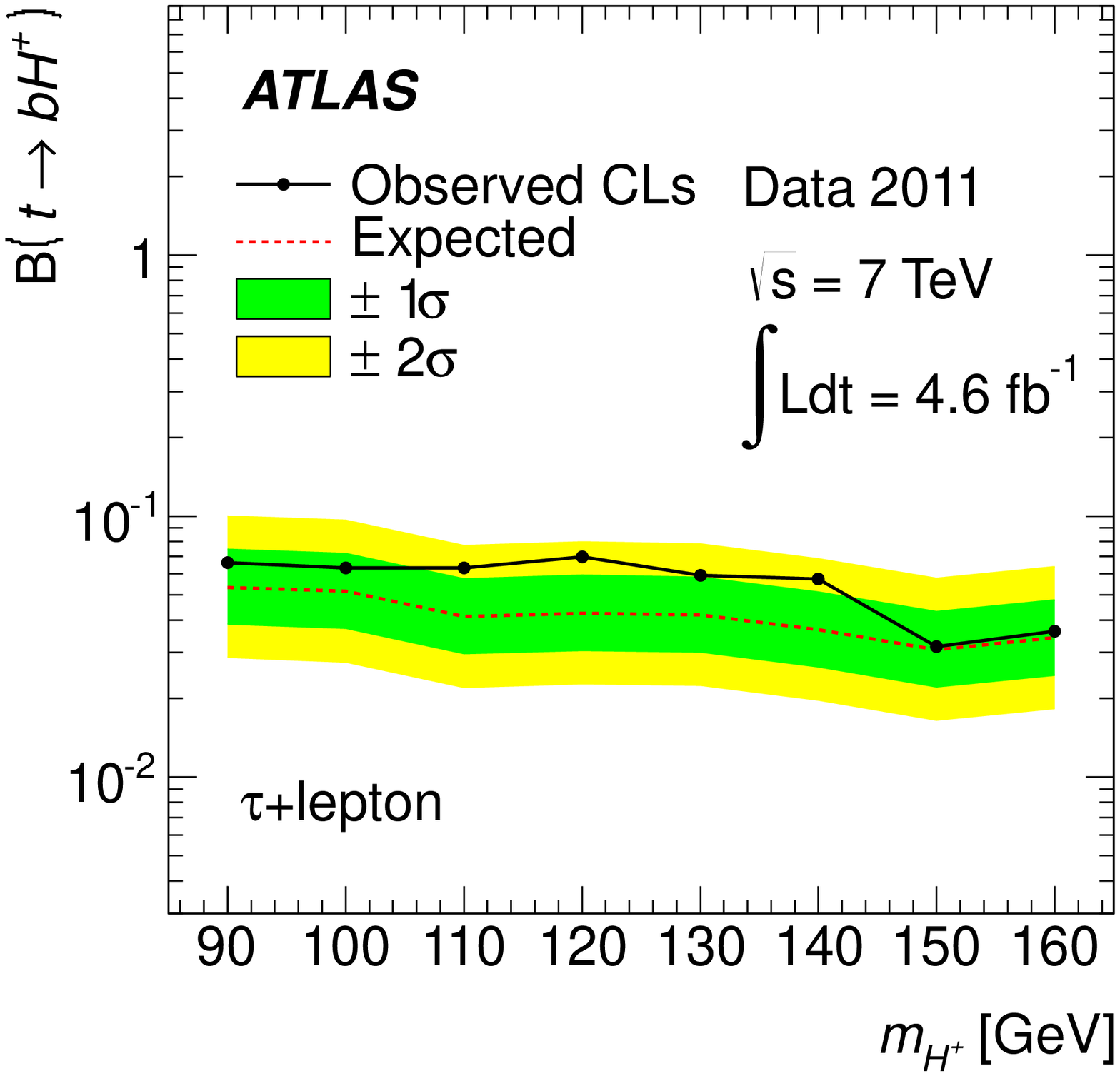}
}
\subfigure[]{
    \includegraphics[width=0.47\textwidth]{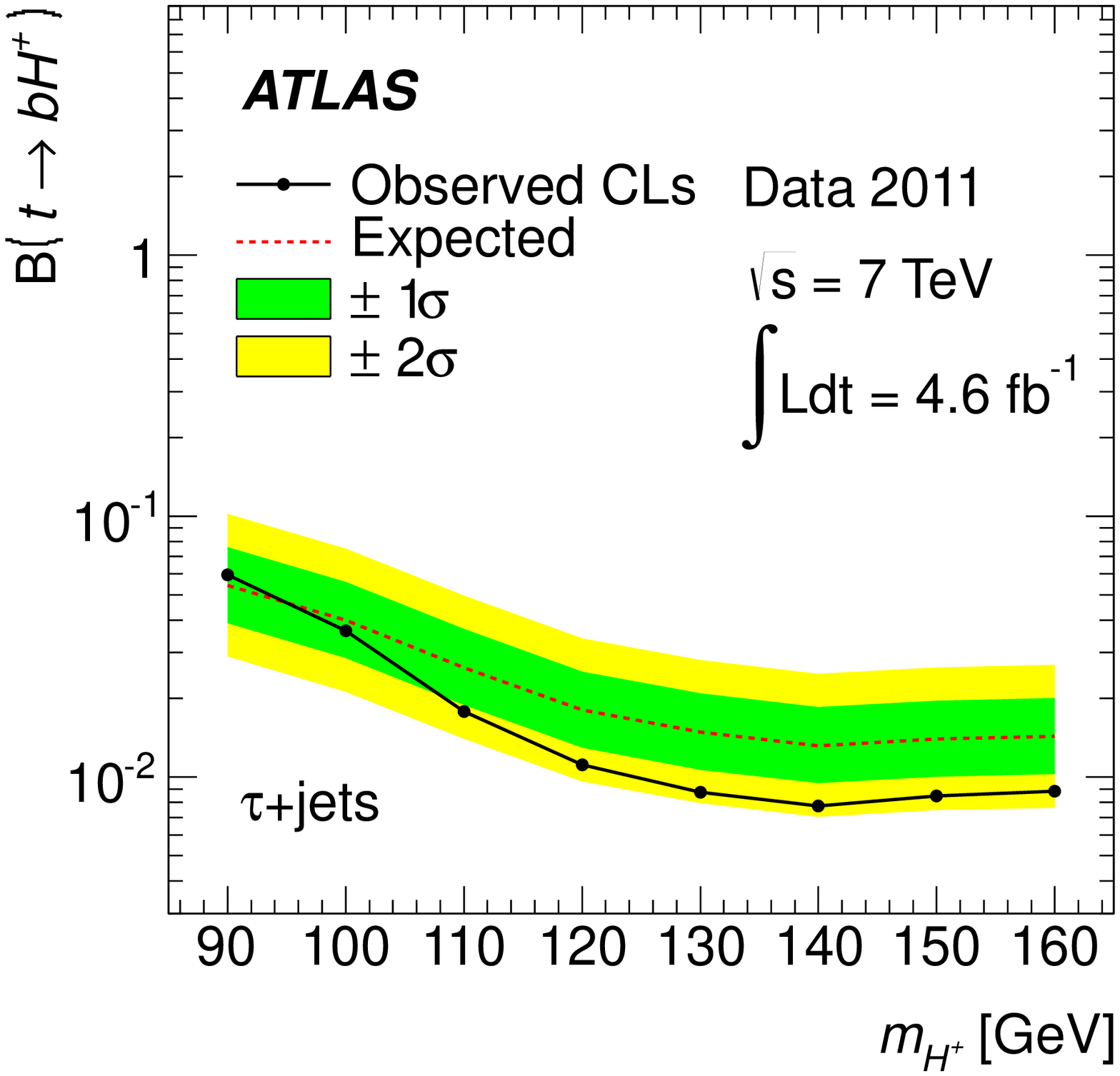}
}
\subfigure[]{
    \includegraphics[width=0.47\textwidth]{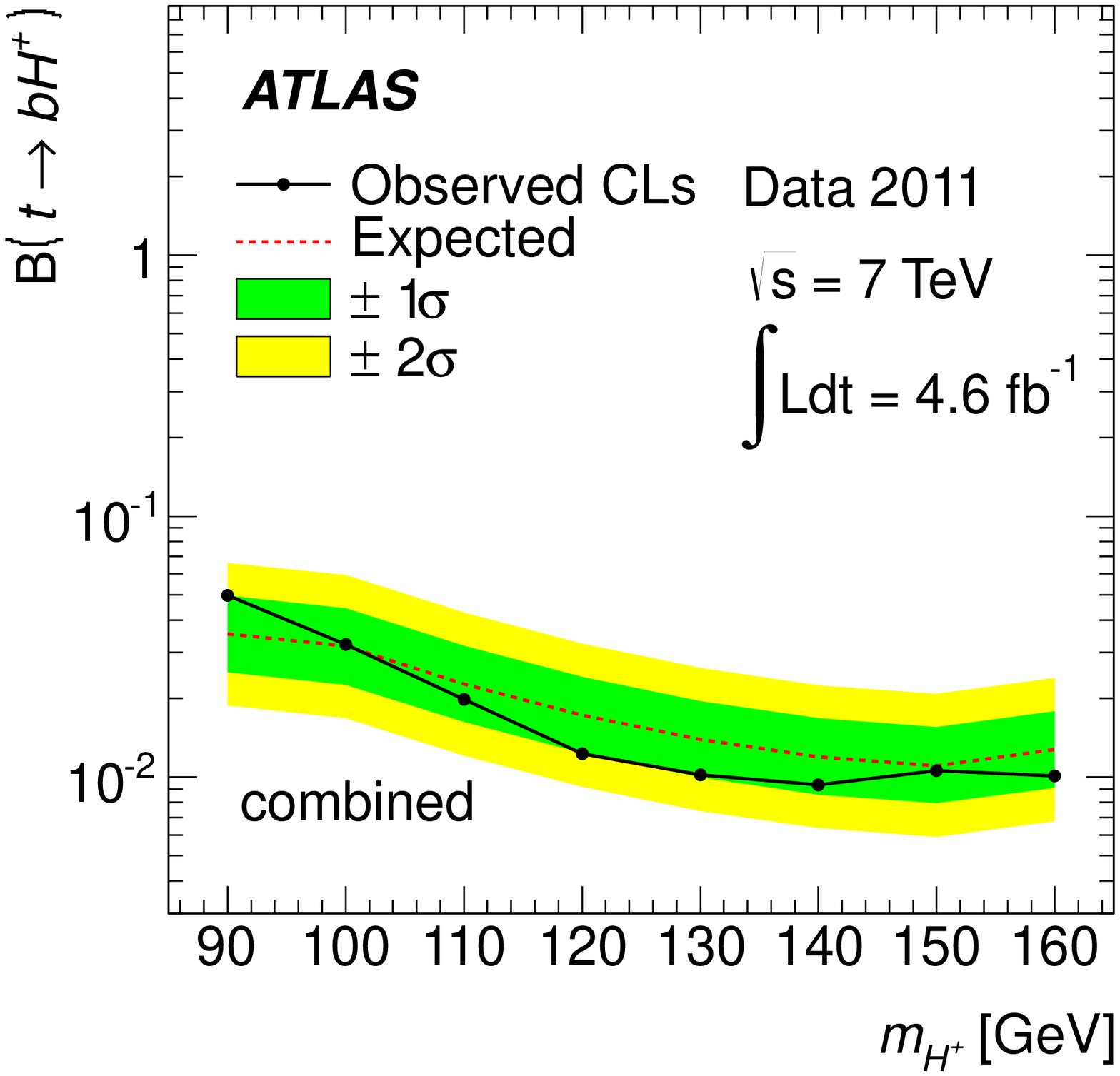}
}
  \caption{Expected and observed 95\% CL exclusion limits on 
${\cal B}(t\rightarrow bH^+)$ for charged 
Higgs boson production from top quark decays as a function of $m_{H^+}$, 
assuming ${\cal B}(H^+\rightarrow\tau\nu)=100\%$. Shown are the results for:
(a) lepton+jets channel; 
(b) $\tau$+lepton channel; (c) $\tau$+jets channel; (d) combination.
}
\label{fig:limit}
\end{figure}

\begin{figure}[t!h!]
  \centering
    \includegraphics[width=0.75\textwidth]{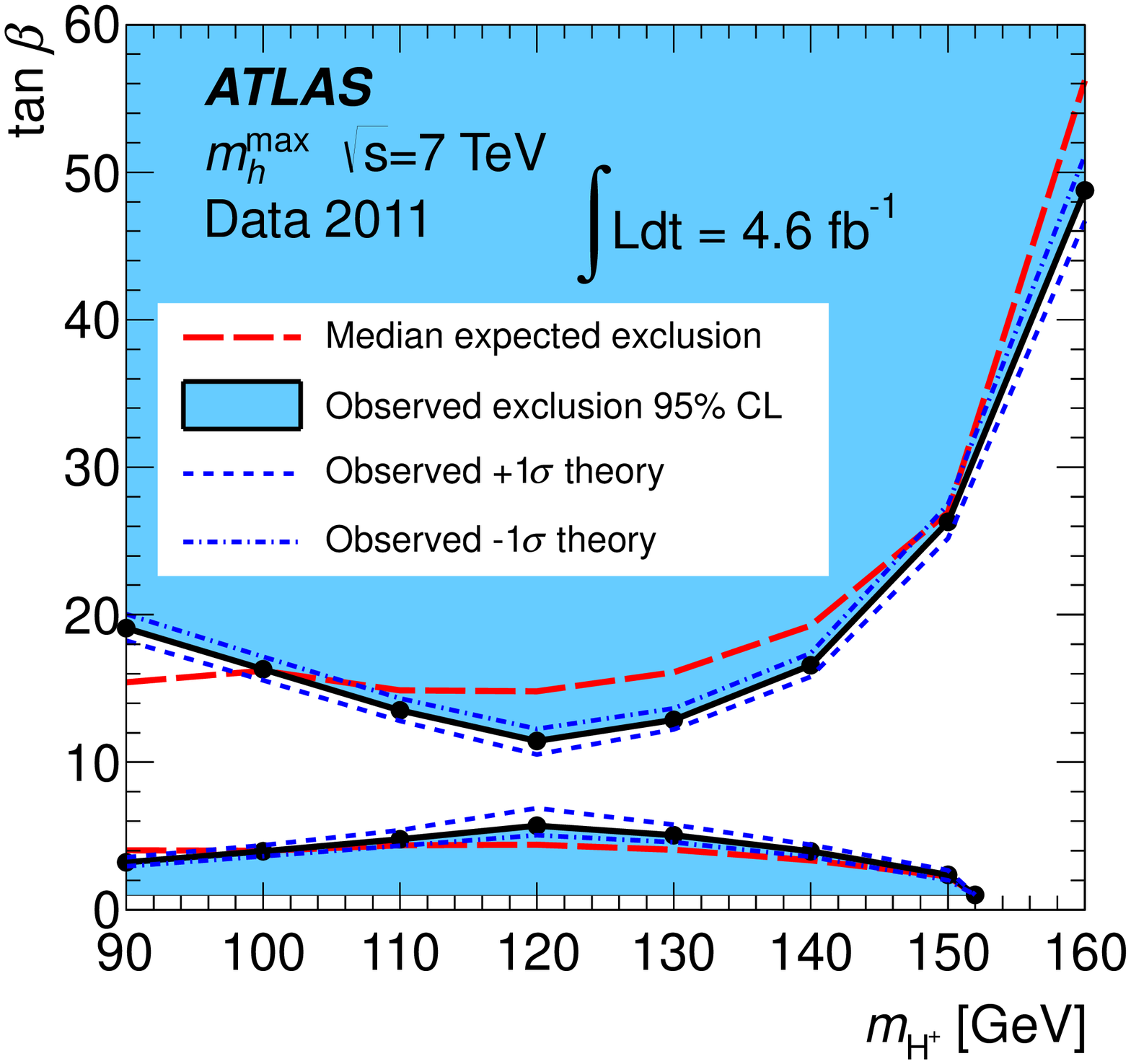}
    \caption{Combined 95\% CL exclusion limits on $\tan\beta$ as a function of 
$m_{H^+}$. Results are shown in the context of the MSSM scenario 
$m_h^{\mathrm{max}}$ for the region $1 < \tan \beta < 60$ in which 
reliable theoretical predictions exist. 
The theoretical uncertainties described in the text are shown as well. 
}
\label{fig:comb_limit_br}
\end{figure}

In order to test the compatibility of the data with background-only 
and signal+background hypotheses, a profile likelihood 
ratio~\cite{Cowan:2010js} is used with $m_{\text{T}}^H$ 
(lepton+jets), $\met$ ($\tau$+lepton) and $m_{\text{T}}$ ($\tau$+jets) 
as the discriminating variables. The statistical analysis is based on a 
binned likelihood function for these distributions. The systematic 
uncertainties 
in shape and normalisation are incorporated via nuisance parameters, and 
the one-sided profile likelihood ratio, $\tilde{q}_{\mu}$, is used as a 
test statistic. No significant deviation from the SM prediction is observed 
in any of the investigated final states in \LUMI of data. 
Exclusion limits are set on the branching fraction ${\cal B}(t \to bH^+)$ 
and, in the context of the $m_h^{\mathrm{max}}$ scenario of the 
MSSM, on $\tan\beta$, by rejecting 
the signal hypothesis at the 95\% confidence level (CL) using the CL$_s$ 
procedure~\cite{Read:2002hq}. These limits are based on the 
asymptotic distribution of the test statistic~\cite{Cowan:2010js}. 
The combined limit is derived from the product of the individual 
likelihoods, and systematic uncertainties are treated as correlated 
where appropriate.
The exclusion limits for the individual channels, as well 
as the combined limit, are shown in Fig.~\ref{fig:limit} in 
terms of ${\cal B}(t\rightarrow bH^+)$ with the assumption 
${\cal B}(H^+\rightarrow\tau\nu)=100\%$. In Fig.~\ref{fig:comb_limit_br}, 
the combined limit on ${\cal B}(t\rightarrow bH^+) \times {\cal B}(H^+\rightarrow\tau\nu)$ 
is interpreted in the context of 
the $m_h^{\mathrm{max}}$ scenario of the MSSM. 
The following relative theoretical uncertainties on 
${\cal B}(t \to bH^+)$ are 
considered~\cite{Heinemeyer:1998yj,Dittmaier:2012vm}: 
5\% for one-loop electroweak corrections missing from the calculations, 
2\% for missing two-loop QCD corrections, and about 1\% (depending on 
$\tan \beta$) for $\Delta_b$-induced uncertainties, where $\Delta_b$ is a 
correction factor to the running $b$ quark mass~\cite{Carena:1999py}. 
These uncertainties are added linearly, as recommended by 
the LHC Higgs cross section working group~\cite{Dittmaier:2012vm}.

\section{Conclusions}
\label{sec:conclusion}

Charged Higgs bosons have been searched for in \ttbar\ events, in the decay 
mode $t \to bH^+$ followed by $H^+ \to \tau\nu$. For this purpose, a total 
of \LUMI of $pp$ collision data at $\sqrt{s}=7~\mbox{TeV}$, recorded in 
2011 with the ATLAS experiment, is used. Three final states are considered, 
which are characterised by the presence 
of a leptonic or hadronic $\tau$ decay, $\met$, $b$ jets, and a leptonically 
or hadronically decaying $W$ boson. Data-driven methods and 
simulation are employed to estimate the number of background events. 
The observed data are found to be in agreement with the SM predictions. 
Assuming ${\cal B}(H^+ \to \tau\nu)=100\%$, 
upper limits at the 95\% confidence level have been set on the 
branching ratio ${\cal B}(t \to bH^{+})$ between 5\% ($m_{H^+}=90\GeV$) 
and 1\% ($m_{H^+}=160\GeV$). This result constitutes a significant 
improvement compared to existing limits provided by the Tevatron 
experiments~\cite{Aaltonen:2009ke,:2009zh} over the whole investigated 
mass range, but in particular for $m_{H^+}$ close to the top quark mass. 
Interpreted in the context of the $m_h^{\rm max}$ scenario of the MSSM, 
$\tan \beta$ above 12--26, as well as between 1 and 2--6, can 
be excluded in the mass range $90 \GeV< m_{H^+}< 150 \GeV$.


\section{Acknowledgements}

We thank CERN for the very successful operation of the LHC, as well as the
support staff from our institutions without whom ATLAS could not be
operated efficiently.

We acknowledge the support of ANPCyT, Argentina; YerPhI, Armenia; ARC,
Australia; BMWF, Austria; ANAS, Azerbaijan; SSTC, Belarus; CNPq and FAPESP,
Brazil; NSERC, NRC and CFI, Canada; CERN; CONICYT, Chile; CAS, MOST and NSFC,
China; COLCIENCIAS, Colombia; MSMT CR, MPO CR and VSC CR, Czech Republic;
DNRF, DNSRC and Lundbeck Foundation, Denmark; EPLANET and ERC, European Union;
IN2P3-CNRS, CEA-DSM/IRFU, France; GNAS, Georgia; BMBF, DFG, HGF, MPG and AvH
Foundation, Germany; GSRT, Greece; ISF, MINERVA, GIF, DIP and Benoziyo Center,
Israel; INFN, Italy; MEXT and JSPS, Japan; CNRST, Morocco; FOM and NWO,
Netherlands; RCN, Norway; MNiSW, Poland; GRICES and FCT, Portugal; MERYS
(MECTS), Romania; MES of Russia and ROSATOM, Russian Federation; JINR; MSTD,
Serbia; MSSR, Slovakia; ARRS and MVZT, Slovenia; DST/NRF, South Africa;
MICINN, Spain; SRC and Wallenberg Foundation, Sweden; SER, SNSF and Cantons of
Bern and Geneva, Switzerland; NSC, Taiwan; TAEK, Turkey; STFC, the Royal
Society and Leverhulme Trust, United Kingdom; DOE and NSF, United States of
America.

The crucial computing support from all WLCG partners is acknowledged
gratefully, in particular from CERN and the ATLAS Tier-1 facilities at
TRIUMF (Canada), NDGF (Denmark, Norway, Sweden), CC-IN2P3 (France),
KIT/GridKA (Germany), INFN-CNAF (Italy), NL-T1 (Netherlands), PIC (Spain),
ASGC (Taiwan), RAL (UK) and BNL (USA) and in the Tier-2 facilities
worldwide.

\bibliographystyle{JHEP}
\bibliography{paper2011}

\clearpage
\begin{flushleft}
{\Large The ATLAS Collaboration}

\bigskip

G.~Aad$^{\rm 48}$,
B.~Abbott$^{\rm 111}$,
J.~Abdallah$^{\rm 11}$,
S.~Abdel~Khalek$^{\rm 115}$,
A.A.~Abdelalim$^{\rm 49}$,
O.~Abdinov$^{\rm 10}$,
B.~Abi$^{\rm 112}$,
M.~Abolins$^{\rm 88}$,
O.S.~AbouZeid$^{\rm 158}$,
H.~Abramowicz$^{\rm 153}$
% ...
\bigskip

% ...
$^{10}$ Institute of Physics, Azerbaijan Academy of Sciences, Baku, Azerbaijan\\
$^{11}$ Institut de F\'isica d'Altes Energies and Departament de F\'isica de la Universitat Aut\`onoma  de Barcelona and ICREA, Barcelona, Spain\\
% ...
$^{48}$ Fakult\"{a}t f\"{u}r Mathematik und Physik, Albert-Ludwigs-Universit\"{a}t, Freiburg i.Br., Germany\\
$^{49}$ Section de Physique, Universit\'e de Gen\`eve, Geneva, Switzerland\\
% ...
$^{88}$ Department of Physics and Astronomy, Michigan State University, East Lansing MI, United States of America\\
% ...
$^{111}$ Homer L. Dodge Department of Physics and Astronomy, University of Oklahoma, Norman OK, United States of America\\
$^{112}$ Department of Physics, Oklahoma State University, Stillwater OK, United States of America\\
% ...
$^{115}$ LAL, Univ. Paris-Sud and CNRS/IN2P3, Orsay, France\\
% ...
$^{153}$ Raymond and Beverly Sackler School of Physics and Astronomy, Tel Aviv University, Tel Aviv, Israel\\
% ...
$^{158}$ Department of Physics, University of Toronto, Toronto ON, Canada\\
% ...
\end{flushleft}

\end{document}